\documentclass[reqno]{amsart}

\usepackage{amsthm,amsmath,amsfonts,amssymb,epsfig,graphicx,latexsym,psfrag,float,subfigure,url}

\newtheorem{definition}{Definition}

\newtheorem{remark}[definition]{Remark}
\newtheorem{algorithm}[definition]{Algorithm}
\newtheorem{assumption}[definition]{Assumption}

\newtheorem{notation}[definition]{Notation}
\newtheorem{model}[definition]{Model}

\begin{document}

\title[Random field sampling for a simplified model of melt-blowing]
{Random field sampling\\ for a simplified model of melt-blowing\\ considering turbulent velocity fluctuations}

\author[F.~H\"ubsch]{Florian H\"ubsch$^{1,2}$}
\author[N.~Marheineke]{Nicole Marheineke$^{3,\star}$}
\author[K.~Ritter]{Klaus Ritter$^2$}
\author[R.~Wegener]{Raimund Wegener$^1$} 

\date{\today\\
$^\star$ \textit{Corresponding author}, \texttt{marheineke@am.uni-erlangen.de}\\
$^1$ Fraunhofer ITWM, Fraunhofer Platz 1, D-67663 Kaiserslautern\\
$^2$ TU Kaiserslautern, FB Mathematik, Computational Stochastics, Paul-Ehrlich-Str.~31, D-67663 Kaiserslautern\\
$^3$ FAU Erlangen-N\"urnberg, Lehrstuhl Angewandte Mathematik I, Cauerstr.~11, D-91058 Erlangen, Germany
}

\begin{abstract} 
In melt-blowing very thin liquid fiber jets are spun due to high-velocity air streams. In literature there is a clear, unsolved discrepancy between the measured and computed jet attenuation. In this paper we will verify numerically that the turbulent velocity fluctuations causing a random aerodynamic drag on the fiber jets -- that has been neglected so far -- are the crucial effect to close this gap. 
For this purpose, we model the velocity fluctuations as vector Gaussian random fields on top of a $\mathrm{k}$-$\epsilon$ turbulence description and develop an efficient sampling procedure. Taking advantage of the special covariance structure the effort of the sampling is linear in the discretization and makes the realization possible. 
\end{abstract}

\maketitle



\section{Introduction}\label{sec:1}

\begin{figure}[t]
\includegraphics[scale=0.6]{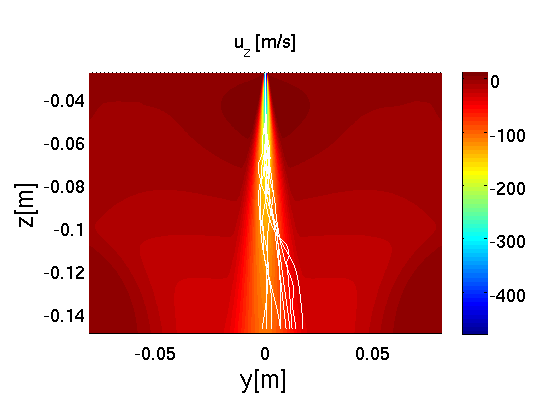}
\vspace*{-0.5cm}
\caption{\label{fig:intro} Simulation of turbulent spinning in a melt-blowing process. Vertically downwards directed high-velocity air stream with immersed spun fiber jets (white curves) whose random motion is caused by the turbulent fluctuations. The color scale visualizes the vertical component of the mean flow velocity. For details see Section~\ref{sec:4}.}
\end{figure}

Melt-blowing is a process for manufacturing very thin thermoplastic fibers, whose commercial importance steadily increases. It dates back to Wente's work in the 1950s at the Naval Research Laboratory in the USA \cite{wente:p:1954}. For an overview on the technology we refer to \cite{pinchuk:b:2002, malkan:p:1995}. 
In a melt-blowing process, a molten stream of polymer is extruded from the spinneret into a forwarding high-velocity air stream. The aerodynamic force rapidly attenuates the polymer jet from a diameter $\mathrm{d}_0$ of approximately 500 micrometers at the nozzle down to final diameters $\mathrm{d}$ that can be as small as 0.5 micrometers. The speeds are very high. Since air and polymer are nearly of the same temperature, the gas prevents polymer solidification at distances close to the die. So fibers are produced that are orders of magnitude smaller than the fibers of a conventional melt-spinning process where the stretching is caused by a mechanical force due to a take-up wheel. The elongation in melt-blowing is $e=\mathrm{A}_0/\mathrm{A}=\mathrm{d}_0^2/\mathrm{d}^2 \sim \mathcal{O}(10^6)$, that means a reduction of $10^3$ in diameter $\mathrm{d}$ and of $10^6$ in cross-sectional area $\mathrm{A}$. Melt-blown fibers make excellent filters. They have a high insulating value, moreover they show high cover, surface area and potentially high strength per unit weight.

The optimization of the fabric and the manufacturing process requires the understanding of the fiber structure development in melt-blowing \cite{bresee:p:2003}. To gain insight in fiber jet attenuation and cooling several on-line measurements have been performed during the last years (see e.g.\ \cite{xie:p:2012, yin:p:1999} and for measurements on jet diameter and temperature \cite{bansal:p:1998, uyttendaele:p:1990}, jet velocity components \cite{wu:p:1992}, frequency and amplitude of jet vibrations \cite{chhabra:p:1996, sinha-ray:p:2010}, nonwoven webs \cite{lee:p:1992} etc.). But, until now there is a clear, unsolved discrepancy between experiments and mathematical models / simulations. The numerical results presented in the literature coincide quite well with the measurements under conditions of a conventional melt-spinning process with moderate elongation $e\sim \mathcal{O}(10^2)$, but absolutely underestimate the jet attenuation in orders of magnitude for industrial melt-blowing processes, \cite{uyttendaele:p:1990, xie:p:2012, zeng:p:2011}. The reason might lie in the fact that the underlying mathematical models have been originally developed for melt-spinning processes, dealing with mass, momentum and energy balances for a steady (longitudinal) spinning threadline, cf.\ first publications \cite{kase:p:1965, matovich:p:1969} in the 1960s or for an overview \cite{ziabicki:b:1985}. Up to now the studies have been extended to viscous and viscoelastic fluids with inclusion of heat transfer, inertial and air drag effects and with regard of jet dynamics, vibrations and bending instabilities. It is an area of active research as recent articles show, see for example \cite{entov:p:1984, marla:p:2003, marheineke:p:2009, marheineke:p:2011, rao:p:1993, sinha-ray:p:2010, sun:p:2011, yarin:b:1993, yarin:p:2010} and references within. However, in the used steady considerations, the jet cross-sectional area $\mathrm{A}$ and the speed $\mathrm{v}$ are related according to $\mathrm{v}_0\mathrm{A}_0=\mathrm{v}\mathrm{A}$ for an incompressible fluid where the index $_0$ indicates the quantities at the nozzle, such that the computed elongation is obviously restricted by the velocity $\mathbf{u}$ of the acting air stream, i.e.\ $e=\mathrm{v}/\mathrm{v}_0<\|\mathbf{u}\|_\infty/\mathrm{v}_0$. This also holds true for advanced melt-blowing simulations with a turbulence model for the high-velocity air stream when only the mean flow field informations are taken into account \cite{zeng:p:2011}. The computed elongation is hence $e\sim \mathcal{O}(10^4)$ -- in contrast to the measured $e\sim \mathcal{O}(10^6)$. Latest experiments \cite{xie:p:2012, sinha-ray:p:2010} indicate the relevance of the turbulent fluctuations for the jet thinning. 

This paper aims at the numerical verification of the crucial effect of the turbulent velocity fluctuations for the fiber jet attenuation (Figure~\ref{fig:intro}). The turbulent fluctuations of the high-velocity air stream turn out to be qualitatively very important, their impact might close the gap between measurements and previous simulations that has been observed and studied for a long time. Since direct numerical simulations (DNS) are computationally not possible, we use a stochastic $\mathrm{k}$-$\epsilon$ turbulence description for the high-velocity air stream, on top we model the turbulent fluctuations as Gaussian random fields in space and time according to \cite{marheineke:p:2006, marheineke:p:2011}. Their covariance/correlation functions satisfy Kolmogorov's 5/3-law for the energy spectrum and the requirements of the $\mathrm{k}$-$\epsilon$ turbulence model. Moreover, they have a special structure, for which we develop an efficient sampling strategy that makes the application possible. Under the conditions of an industrial melt-blowing process we demonstrate the relevance of the turbulent fluctuations and the resulting random aerodynamic drag force in a one-way-coupling between the air stream and the fiber jet. Thereby, we use an isothermal model for the fiber jet dynamics that consists of a system of ordinary differential equations (ODE) for jet position, velocity and elongation. Already for this very simplified ODE-model with randomness we obtain a jet thinning behavior that is qualitatively appropriate in magnitude. The results are very promising and raise hope that the application of the random aerodynamic drag to more sophisticated Cosserat models for the jet dynamics \cite{arne:p:2010, arne:p:2011, ribe:p:2004, ribe:p:2006} that also include inner stresses and temperature dependencies (systems of partial differential equations (PDE)) will finally answer the open questions of the fiber structure development in melt-blowing in future. Recent work deals with the robust numerical treatment of the PDE-models \cite{arne:p:2012, audoly:p:2012}. Of particular importance is thereby the establishment of a fast and accurate adaptive mesh refinement that is necessary in order to cope with the expected huge elongations.

The paper is structured as follows. After a short survey in turbulence modeling, the reconstruction of the turbulent velocity fluctuations as Gaussian random fields taken from \cite{marheineke:p:2006} is presented in Section~\ref{sec:2}. In Section~\ref{sec:3} we develop a suitable, very efficient sampling strategy that makes use of the special covariance structure. We apply the resulting random aerodynamic drag force to a simplified ODE-model for melt-blowing in Section~\ref{sec:4}. Comparing the results with regard and neglect of the turbulent fluctuations clearly shows the importance of the fluctuations for a proper description of the manufacturing process.

\section{Turbulence Modeling}\label{sec:2}

All the physics of a turbulent flow is contained in the non-stationary Navier-Stokes equations (NSE). Turbulence can be considered as continuum phenomenon, since even the smallest scales occurring in a turbulent flow are ordinarily far larger than any molecular scales. Thus, solving NSE by means of DNS gives the exact velocity field needed for the determination of the aerodynamic force causing the jet stretching. However, DNS presupposes the resolution of all vortices ranging from the large energy-bearing
ones of length $\mathrm{l_T}$ to the smallest, viscously determined Kolmogorov vortices of size $\eta$ with $\mathrm{l}_{\mathrm{T}}/\eta=\mathrm{Re}^{3/4}$ \cite{wilcox:b:1998}. Hence, the number of grid points that are required for the refinement of a three-dimensional domain is proportional to $\mathrm{Re}^{9/4}$. Despite of recent high speed computations, DNS is thus still restricted to
simple, small Reynolds number flow. In large eddy simulations (LES) the computational effort is reduced. By applying a low-pass filter on NSE, only the vortices of large scales are resolved directly, whereas the small vortices are taken into account by a stochastic approximation of their effect on the larger ones. This procedure works well for fluid-structure interactions with structures of moderate size, e.g.\ flow around the wing of an aircraft. However, in view of our thin fiber jets we need correlation information and characteristics of such small scales so that LES is not applicable. Therefore, we use stochastic turbulence models. These models suffer from their deficient description of boundary layers at walls and/or obstacles. But since we restrict to a one-way-coupling by neglecting the effect of the few slender jets on the air flow, the flow information that we need for the air drag model comes from the simulation of a turbulent flow in a machine geometry without any immersed fiber jets. So the vortex structure develops in the inner flow domain, unperturbed by any obstacles. And no inaccurate boundary layer approximations falsify the numerical results. The stochastic turbulence models -- also known as statistical turbulence models \cite{ferziger:b:2002, wilcox:b:1998} -- prescribe the instantaneous space- and time-dependent velocity field $\mathbf{u}:\mathbb{R}^3 \times \mathbb{R}_0^+\rightarrow \mathbb{R}^3$ as sum of a mean (deterministic) $\mathbf{\bar u}$ and a fluctuating (stochastic) part $\mathbf{u'}$
\begin{align*}
\mathbf{u}=\mathbf{\bar u} + \mathbf{u'}.
\end{align*}
So, the velocity field is considered as a $\mathbb{R}^3$-valued random field $\mathbf{u}=(\mathbf{u}(\mathbf{x},\mathrm{t}))_{(\mathbf{x},\mathrm{t})\in \mathbb{R}^3\times \mathbb{R}^+_0}$, i.e., the velocity at the point $(\mathbf{x},\mathrm{t})$ in space and time is assumed to be a realization (sample) of the $\mathbb{R}^3$-valued random variable (random vector) $\mathbf{u}(\mathbf{x},\mathrm{t})$. On the average over all realizations $\mathbf{u}$ equals $\mathbf{\bar u}$, i.e., $\mathbb{E}(\mathbf{u})=\mathbf{\bar u}$ or, equivalently $\mathbb{E}(\mathbf{u'})=\mathbf{0}$, where $\mathbb{E}$ denotes the expectation. The mean velocity $\mathbf{\bar u}$ is determined by the Reynolds-averaged Navier-Stokes equations (RANS) where the effect of the fluctuations is contained as source in terms of the Reynolds stress tensor $\mathbf{T}=-\mathbb{E}(\mathbf{u'}\otimes \mathbf{u'})$ in the momentum balance. The fluctuations $\mathbf{u'}=(\mathbf{u'}(\mathbf{x},\mathrm{t}))_{(\mathbf{x},\mathrm{t})\in \mathbb{R}^3\times \mathbb{R}^+_0}$ are not modeled directly themselves as centered random field, instead the stochastic turbulence models restrict on modeling the symmetric Reynolds stress tensor. Therefore additional statistical quantities are introduced. In the $\mathrm{k}$-$\epsilon$ model, one of the prime turbulence models going back to Launder and Sharma \cite{launder:p:1974}, these quantities are the turbulent kinetic energy $\mathrm{k}:\mathbb{R}^3\times \mathbb{R}_0^+ \rightarrow \mathbb{R}^+$ and the dissipation rate $\epsilon:\mathbb{R}^3\times \mathbb{R}_0^+\rightarrow \mathbb{R}^+$. They characterize the fluctuations 
\begin{align}\label{eq:keps}
\mathrm{k}&=\frac{1}{2} \,\mathbb{E}(\mathbf{u'}\cdot \mathbf{u'}), \quad \quad \quad
\epsilon=\nu\,\mathbb{E}(\nabla \mathbf{u'}: \nabla \mathbf{u'})
\end{align}
with flow viscosity $\nu$ and yield the Reynolds stress tensor $\mathbf{T}=C_\mu \mathrm{k}^2/ \epsilon \,(\nabla \mathbf{\bar u} +\nabla \mathbf{\bar u}^T) -2\mathrm{k}/3 \boldsymbol{\mathit{I}}$ with unit tensor $\boldsymbol{\mathit{I}}$ and constant $C_\mu$ in accordance with the Boussinesq assumption that proceeds from the analogy between viscous and turbulent stresses. In general, the purely deterministic PDE-models for the turbulence description consist of RANS and transport equations for the additional statistical quantities. They are closed by means of numerous assumptions and fitted parameters (closure relations).

\begin{notation}[Tensor Calculus]
A tensor can be associated with a multi-dimensional array with regard to the chosen basis, e.g.\ a tensor of 2nd, 1st or 0th order is represented by a matrix, vector or scalar, respectively.  Throughout the paper we use large and small bold-faced letters for matrix- and vector-valued quantities, respectively. Scalar-valued quantities are typeset in normal-faced letters. Elementary operations include tensor and dot products, such as $\mathbf{a}\otimes \mathbf{b}=\mathbf{C}$ with $\mathrm{C}_{ij}=\mathrm{a}_i\mathrm{b}_j$ as well as $\mathbf{a}\cdot \mathbf{b}= \sum_i \mathrm{a}_i \mathrm{b}_i$ and $\mathbf{A}:\mathbf{B}=\sum_{i,j} \mathrm{A}_{ij} \mathrm{B}_{ij}$ (scalar products of vectors and matrices).
\end{notation}

On top of a $\mathrm{k}$-$\epsilon$ formulation we developed a turbulence reconstruction strategy for $\mathbf{u}'$ in \cite{marheineke:p:2006}. This strategy is based on a Global-from-Local-Assumption according to that the local velocity fluctuations (fine-scale structure) are modeled as homogenous, isotropic Gaussian random fields that are superposed to form the large-scale structure of the global turbulence. This assumption is motivated by Kolmogorov's local isotropy hypothesis \cite{frisch:b:1995}: certain theoretical considerations concerning the energy transfer through the eddy-size spectrum from the larger to the smaller eddies lead to the conclusion that the fine structure of anisotropic turbulent flows is almost isotropic.
\begin{assumption}[Global-from-Local-Assumption \cite{marheineke:p:2006}] \label{ass:1}
Let the $\mathrm{k}$-$\epsilon$ turbulence model be given. Let $(\Omega,\mathcal{A},\mathrm{P})$ be a probability space.
Let $\mathbf{u'}_{loc,\mathrm{p}}=(\mathbf{u'}_{loc,\mathrm{p}}(\mathbf{x},\mathrm{t}))_{(\mathbf{x},\mathrm{t})\in \mathbb{R}^3\times
\mathbb{R}_0^+}$ be a family of centered Gaussian random fields on $(\Omega,\mathcal{A},\mathrm{P})$ that correspond to spatially and temporally homogeneous, isotropic, and incompressible flow fluctuations with respect to the local turbulence information ($\mathrm{k}$, $\epsilon$, $\nu$, $\mathbf{\bar u}$) at the point $\mathrm{p}=(\mathbf{y},\tau)\in \mathbb{R}^3\times \mathbb{R}^+_0$. Their tensor-valued covariance/correlation functions are denoted by $\mathbf{K}_{loc,\mathrm{p}}:(\mathbb{R}^3\times\mathbb{R}_0^+)^2\rightarrow \mathbb{R}^{3\times3}$.

Then we assume that the actual global fluctuation field $\mathbf{u}'$ can be constructed as
\begin{align}\label{eq:u'}
\mathbf{u'}(\mathbf{x},\mathrm{t}) =\langle \mathbf{u'}_{loc,\mathrm{p}}(\mathbf{x},\mathrm{t})\rangle_{M(\mathbf{x},\mathrm{t})},
\end{align}
with $M(\mathbf{x},\mathrm{t})=\{\mathrm{p}=(\mathbf{y},\tau)\in \mathbb{R}^3\times \mathbb{R}^+_0\,|\, 
  \|\mathbf{x}-\mathbf{y}-\mathbf{\bar u}(\mathbf{x},\mathrm{t}) (\mathrm{t}-\tau)\|_2 \leq \mathrm{l}_\mathrm{T} \,\wedge \,
  |\mathrm{t}-\tau|\leq \mathrm{t}_\mathrm{T}\} $, $|{M}(\mathbf{x},\mathrm{t})|=\int_{{M}(\mathbf{x},\mathrm{t})} \mathrm{d}\mathrm{p}$, 
  and turbulent large-scale length $\mathrm{l}_{\mathrm{T}}$ and time $\mathrm{t}_{\mathrm{T}}$.
The brackets $\langle \cdot\rangle$ stand for a Gaussian average with respect to the parameter $\mathrm{p}$, i.e., $\mathbf{u}'$ is a Gaussian random field that is uniquely prescribed by the following mean and covariance function
\begin{align*}
\mathbb{E}(\mathbf{u'}(\mathbf{x},\mathrm{t}))&=
\frac{1}{|M(\mathbf{x},\mathrm{t})|}\int_{M(\mathbf{x},\mathrm{t})} \mathbb{E}(\mathbf{u'}_{loc,\mathrm{p}}(\mathbf{x},\mathrm{t}))\,\mathrm{d}\mathrm{p}=\mathbf{0}, \\
\mathbb{E}(\mathbf{u'}(\mathbf{x},\mathrm{t})\otimes\mathbf{u'}(\mathbf{x_1},\mathrm{t}_1))&= 
\frac{1}{\sqrt{|M(\mathbf{x},\mathrm{t})||M(\mathbf{x_1},\mathrm{t}_1)|}} \int_{M(\mathbf{x},\mathrm{t})\cap M(\mathbf{x_1},\mathrm{t}_1)}
 \hspace*{-1cm} \mathbf{K}_{loc,\mathrm{p}}(\mathbf{x},\mathrm{t},\mathbf{x_1},\mathrm{t}_1) \,\mathrm{d}\mathrm{p}\\
&=\mathbf{K}(\mathbf{x},\mathrm{t},\mathbf{x_1},\mathrm{t}_1).
\end{align*} 
\end{assumption}

\begin{remark}
A random field is Gaussian if all its finite-dimensional joint distributions are Gaussian (normal distributed). The distribution of a Gaussian random field is completely specified by its mean and covariance function. By Assumption~\ref{ass:1} the modeling of the turbulent fluctuations $\mathbf{u}'$ is reduced to the modeling of $\mathbf{u}'_{loc,\mathrm{p}}$ focusing on an appropriate description of the covariance function $\mathbf{K}_{loc,\mathrm{p}}$. The covariance function contains the information about the spatial and temporal correlations in the turbulent flow.
\end{remark}

The local centered Gaussian velocity fluctuation fields $\mathbf{u'}_{loc,\mathrm{p}}$ and hence their covariance/correla\-tions $\mathbf{K}_{loc,\mathrm{p}}$ depend parametrically on the flow situation $(\mathrm{k},\epsilon,\nu,\mathbf{\bar u})$ at the point $\mathrm{p}$ that is provided by the $\mathrm{k}$-$\epsilon$ model. Thus, we make $\mathbf{u'}_{loc,\mathrm{p}}$ dimensionless using the typical turbulent length $\mathrm{k}^{3/2}/\epsilon$ and time $\mathrm{k}/\epsilon$ corresponding to $\mathrm{p}$, 
\begin{align}\label{eq:u}
 \mathbf{u'}_{loc}(\mathbf{x},\mathrm{t})=\mathrm{k}^{1/2} \,\boldsymbol{\mathit{u'}}_{loc}\left(\frac{\epsilon}{\mathrm{k}^{3/2}}\mathbf{x},\frac{\epsilon}{\mathrm{k}}\mathrm{t}; \frac{\epsilon}{\mathrm{k}^2}\nu \right), \quad \quad \mathbf{x}=\frac{\mathrm{k}^{3/2}}{\epsilon}\boldsymbol{\mathit{x}}, \quad \mathrm{t}=\frac{\mathrm{k}}{\epsilon}t, \quad \nu=\frac{\mathrm{k}^2}{\epsilon}\zeta.
\end{align}
The dimensionless viscosity $\zeta$ enters the model of the velocity fluctuations via the consistency with the $\mathrm{k}$-$\epsilon$ description, see Model~\ref{mod:E}. To facilitate the readability we suppress the parameter-dependence (index $\mathrm{p}$) here in \eqref{eq:u} and also in the following. 

\begin{notation}[Dimensional vs Dimensionless Quantity]
We typeset dimensional quantities in Roman style (e.g.\ $\mathrm{t}$, $\mathbf{x}$, $\mathbf{u'}$, $\mathbf{K}$) and the corresponding dimensionless quantities in Italic style (e.g.\ $t$, $\boldsymbol{\mathit{x}}$, $\boldsymbol{\mathit{u'}}$, $\boldsymbol{\mathit{K}}$) throughout the paper.
\end{notation}

In the forthcoming explanations we focus on the dimensionless quantities.
The development of the local tensor-valued covariance/correlations $\boldsymbol{\mathit{K}}_{loc}$ can be reduced to the modeling of two scalar-valued functions, the energy spectrum $E$ and the decay of the temporal correlations $\varphi$. (For the detailed derivation of the original model with frozen turbulence pattern we refer to \cite{marheineke:p:2006}, extensions on the temporal correlations are studied and incorporated in \cite{marheineke:p:2011}. For more informations about turbulence and its evolution see \cite{frisch:b:1995, kaneda:p:1993, majda:p:1999, marheineke:p:2011} and references within.) 
In a homogeneous turbulent flow, the correlations are invariant with regard to spatial and temporal translations and hence depend only on the differences of the arguments. The evolution of the correlations are modeled by an advection-driven vortex structure that is naturally decaying over time (alleviated frozen turbulence).  Taylor's hypothesis of frozen turbulence pattern \cite{taylor:p:1938}, i.e.\ fluctuations arise due to so-called turbulence pattern that are transported by the mean flow without changing their structure, is based on the observation that the rate of decay of the mean properties is rather slow with respect to the time scale of the fluctuating fine-scale structures. The superposition with a natural temporal decay is essential for describing suspensions of particles or filaments in turbulent flows, since otherwise small light objects tending to move with the mean flow field would experience permanently the same non-varying fluctuations. So, $\boldsymbol{\mathit{K}}_{loc}$ is prescribed by the initial correlation tensor $\boldsymbol{\gamma}:\mathbb{R}^3\rightarrow \mathbb{R}^{3\times3}$ and the temporal decay function $\varphi:\mathbb{R}_0^+\rightarrow \mathbb{R}$, 
\begin{align}\label{eq:Kloc}
\boldsymbol{\mathit{K}}_{loc}(\boldsymbol{\mathit{x}}+\boldsymbol{\mathit{x_1}},t+t_1,\boldsymbol{\mathit{x_1}},t_1)= \boldsymbol{\gamma}(\boldsymbol{\mathit{x}}-\boldsymbol{\mathit{\bar u}}t)\,\, \varphi(t) 
\end{align}
with mean flow velocity $\boldsymbol{\mathit{\bar u}}$. It is also made dimensionless with respect to the typical turbulent length and time in consistency to \eqref{eq:u}. In case of incompressible isotropic turbulence, $\boldsymbol{\gamma}$ is represented by a single scalar-valued smooth function. In terms of the spectral density being its Fourier transform $\mathcal{F}_{\boldsymbol{\gamma}}$, this function is known as energy spectrum $E:\mathbb{R}_0^+\rightarrow \mathbb{R}_0^+$ that has been well-studied theoretically and experimentally in the last century,
\begin{align}\label{eq:Sloc}
\mathcal{F}_{\boldsymbol{\gamma}}(\boldsymbol{\kappa})=
\frac{1}{(2\pi)^{3}} \int_{\mathbb{R}^3} \exp(-i\boldsymbol{\kappa}\cdot \boldsymbol{\mathit{x}})\,\boldsymbol{\gamma}(\boldsymbol{\mathit{x}}) \, 
\mathrm{d} \boldsymbol{\mathit{x}} 
= \frac{1}{4\pi} \frac{E(\|\boldsymbol{\kappa}\|)}{\|\boldsymbol{\kappa}\|^2}
\left(\boldsymbol{\mathit{I}}-\frac{1}{\|\boldsymbol{\kappa}\|^2} \boldsymbol{\kappa}\otimes \boldsymbol{\kappa}\right)
\end{align}
with unit tensor $\boldsymbol{\mathit{I}}$ and Euclidian norm $\|.\|$.
Kolmogorov's universal equilibrium theory was thereby trend setting. Based on dimensional analysis he derived not only the characteristic ranges but also the typical behavior of the spectrum which agrees with later coming physical concepts and experiments, cf.\ Kolmogorov's 5/3-law and his hypothesis of local isotropy \cite{frisch:b:1995}. Gathering the existing knowledge about $E$, an appropriate model has to satisfy  Kolmogorov's 5/3-law as well as the requirements of the $\mathrm{k}$-$\epsilon$ turbulence model, i.e. 
\begin{align*}
\int_0^\infty E(\kappa) \,\mathrm{d}\kappa=1, \quad \quad
\int_0^\infty \kappa^2 E(\kappa) \, \mathrm{d}\kappa= \frac{1}{2\zeta}, \quad \quad
\int_0^\infty (\ln(1+\kappa))^\alpha \kappa^2 E(\kappa) \,\mathrm{d}\kappa< \infty \text{ for some } \alpha>3.
\end{align*}
The first two relations correspond to the definitions of the kinetic turbulent energy $\mathrm{k}$ and dissipation rate $\epsilon$ in \eqref{eq:keps}. For $\epsilon$ to make sense, the third relation ensures that the fluctuation field is almost surely sample differentiable in space. (The $n$-times sample differentiability of a Gaussian field results from certain integral properties of its spectral (density) function, for details see \cite{kruse:d:2001}.) The conditions on $E$ allow for a family of functions that can be adapted to experiments. The parametric $\zeta$-dependence comes from non-dimensionalizing \eqref{eq:keps} and is handed over to the correlation tensor and $\boldsymbol{\mathit{u'}}_{loc}$, cf.\ \eqref{eq:u}.
 
\begin{model}[Energy Spectrum \cite{marheineke:p:2006, marheineke:p:2011}]\label{mod:E}
The energy spectrum is modeled as $E\in \mathcal{C}^2(\mathbb{R}^+_0)$
\begin{align}\label{eq:E}
  E(\kappa;\zeta) &= C_{\mathrm K} 
\left \{\begin{array}{ll}
\kappa_1^{-5/3} \,\,\sum_{j=4}^6 \,a_j\, (\frac{\kappa}{\kappa_1})^j & \quad \kappa < \kappa_1\\
\kappa^{-5/3} & \quad \kappa_1 \leq \kappa \leq \kappa_2\\
\kappa_2^{-5/3} \,\,\sum_{j=7}^9 \,b_j\, (\frac{\kappa}{\kappa_2})^{-j}  & \quad \kappa_2 < \kappa
\end{array} \right. 
\end{align}
where the $\zeta$-dependent transition wave numbers $\kappa_1$ and $\kappa_2$ are implicitly given by
\begin{align*}
\int_0^\infty E&(\kappa;\zeta) \, \mathrm{d} \kappa = 1, \quad \quad \quad
\int_0^\infty \kappa^2 \, E(\kappa;\zeta) \,\mathrm{d} \kappa = \frac{1}{2\zeta}.
\end{align*}
The regularity parameters are $a_4=230/9$, $a_5=-391/9$, $a_6=170/9$, $b_7=209/9$, $b_8=-352/9$, $b_9=152/9$, and the Kolmogorov constant is $C_{\mathrm{K}}=1/2$.
\end{model}
The integral conditions for $\kappa_1$ and $\kappa_2$ in $\zeta$ \eqref{eq:E} can be reformulated as nonlinear system
\begin{align} \label{eq:E_NGS}
\hat{a}_1 \kappa_1^{-2/3}-\hat{b}_1\kappa_2^{-2/3}=C_K^{-1}, \quad \quad &\quad 
-\hat{a}_2 \kappa_1^{4/3}+\hat{b}_2\kappa_2^{4/3}=(2C_K\zeta)^{-1},\\\nonumber
\hat{a}_1=\frac{3}{2}+\sum_{j=4}^6 \frac{a_j}{j+1},\quad \hat{a}_2=\frac{3}{4}-\sum_{j=4}^6 \frac{a_j}{j+3},\quad &\hat{b}_1=\frac{3}{2}-\sum_{j=7}^9 \frac{b_j}{j-1},\quad \hat{b}_2=\frac{3}{4}+\sum_{j=7}^9 \frac{b_j}{j-3}.
\end{align}
The condition $0<\kappa_1<\kappa_2<\infty$ is equivalent to $0<\zeta<\zeta_{crit}=(2C_K^3(\hat b_2-\hat a_2)(\hat b_1-\hat a_1)^2)^{-1}\approx3.86$.
The bounds on $\zeta$ (where we have $\kappa_1=\kappa_2=(C_K(\hat a_1-\hat b_1))^{3/2}$ for $\zeta=\zeta_{crit}$ and $\kappa_1=(C_{\mathrm{K}}\hat a_1)^{3/2}$, $\kappa_2=\infty$ for $\zeta=0$) are no practically relevant restrictions, since the general turbulence theory assumes the ratio of fine-scale and large-scale length to satisfy $\zeta=\epsilon \nu/\mathrm{k}^2\ll 1$. 

The temporal correlation function $\varphi$ satisfies $\varphi(0)=1$ which implies that the integral of its Fourier transform $\mathcal{F}_\varphi$ is normalized. We use an exponential decay with respect to the turbulent large-scale time with $t_{\mathrm{T}}=0.212$, see e.g.\ \cite{lu:p:1995,pismen:p:1978} and references within.

\begin{model}[Temporal Correlations]
The natural decay of the temporal correlations is modeled as $\varphi\in \mathcal{C}^\infty(\mathbb{R}_0^+)$  
\begin{align} \label{eq:varphi}
\varphi(t)=\exp\left(\frac{-t^2}{2 t^2_{\mathrm{T}}}\right), \quad \quad \mathcal{F}_\varphi(\omega)=\frac{t_{\mathrm{T}}}{\sqrt{2\pi}}\exp\left(\frac{-t_{\mathrm{T}}^2\omega^2}{2}\right),\quad \quad t_{\mathrm{T}}=0.212.
\end{align}
\end{model}

\begin{remark}
The correlated global random field $\mathbf{u'}$ \eqref{eq:u'} can be asymptotically reduced to Gaussian white noise with flow-dependent amplitude. For certain applications this simplification is qualitatively and quantitatively justified and implies an enormous reduction of computational time and memory (see \cite{marheineke:p:2006} for a theoretical localization strategy and the strict asymptotic derivation as well as \cite{marheineke:p:2011} for the application and experimental validation in a production process of nonwoven materials).
\end{remark}

\section{Sampling of Gaussian Random Fields}\label{sec:3}

In this section we deal with the reconstruction and simulation of the local centered Gaussian random velocity fields $\boldsymbol{\mathit{u'}}_{loc}$.  
In view of the prescribed data there exist various reconstruction/sampling procedures in literature, e.g.\ Karhunen-Loeve expansion, Cholesky decomposition, circulant embedding for a given covariance function \cite{cameron:p:2003} or spectral, Fourier, Fourier-wavelet methods for a given spectral function \cite{elliott:p:1997, horntrop:p:1997, kurbanmuradov:p:2006}; for an overview see \cite{asmussen:b:2007, kramer:p:2001} and references within. We propose a technique that takes advantage of the special structure of the given data and turns out to be exact in the covariance and very efficient as we will comment on. Note that in the following we assume the existence of all occurring random fields and stochastic processes as we construct them later on.

\subsection{Construction of $\boldsymbol{\mathit{u'}}_{loc}$}
Considering the covariance function \eqref{eq:Kloc} of the local velocity field, we separate the spatial and temporal arguments by introducing a new Gaussian random field $\boldsymbol{\eta}=(\boldsymbol{\eta}(\boldsymbol{\mathit{x}},t))_{(\boldsymbol{\mathit{x}},t)\in \mathbb{R}^4}$
\begin{align*}
\boldsymbol{\eta}(\boldsymbol{\mathit{x}},t) &= \boldsymbol{\mathit{u'}}_{loc}(\boldsymbol{\mathit{x}} + \boldsymbol{\mathit{\bar u}}t,t)
\end{align*}
from which $\boldsymbol{\mathit{u'}}_{loc}$ can be easily regained.
Its covariance satisfies 
\begin{align}\label{eq:cov_eta}
 \mathbb{E}(\boldsymbol{\eta}(\boldsymbol{\mathit{x}}+\boldsymbol{\mathit{x_1}},t+t_1)\otimes \boldsymbol{\eta}(\boldsymbol{\mathit{x_1}},t_1))=\boldsymbol{\gamma}(\boldsymbol{\mathit{x}})\,\varphi(t).
\end{align}
Let the vector random field $\boldsymbol{\xi} =(\boldsymbol{\xi}(\boldsymbol{\mathit{x}}))_{\boldsymbol{\mathit{x}}\in\mathbb{R}^3}$ and the scalar stochastic process $\psi = (\psi(t))_{t\in\mathbb{R}}$ be centered and stochastically independent with covariance functions 
\begin{align*}
\mathbb{E}\left(\boldsymbol{\xi}(\boldsymbol{\mathit{x}}+\boldsymbol{\mathit{x_1}})\otimes \boldsymbol{\xi}(\boldsymbol{\mathit{x_1}})\right) 
= \boldsymbol{\gamma}(\boldsymbol{\mathit{x}}), \quad \quad \quad
\mathbb{E}\left({\psi}(t+t_1) {\psi}(t_1)\right) = \varphi(t).
\end{align*}
Defining a random field $\tilde{\boldsymbol{\eta}}$ by
\begin{align*}
\tilde{\boldsymbol{\eta}}(\boldsymbol{\mathit{x}},t) = \boldsymbol{\xi}(\boldsymbol{\mathit{x}})\,\psi(t),
\end{align*}
$\tilde{\boldsymbol{\eta}}$ and $\boldsymbol{\eta}$ possess the same covariance function \eqref{eq:cov_eta}. As we are interested in a Gaussian field, we consider
\begin{align}\label{eq:Gauss}
\tilde{\boldsymbol{\eta}}_N(\boldsymbol{\mathit{x}},t) = \frac{1}{\sqrt{N}} \sum_{l=1}^N \tilde{\boldsymbol{\eta}}^{(l)}(\boldsymbol{\mathit{x}},t), \quad \quad N\in \mathbb{N},
\end{align}
in which $\tilde{\boldsymbol{\eta}}^{(l)}$, $l=1,...,N$ are independent copies of $\tilde{\boldsymbol{\eta}}$. The central limit theorem ensures then the convergence in distribution\\
\begin{align*}
 \tilde{\boldsymbol{\eta}}_N(\boldsymbol{\mathit{x}},t) \stackrel{d}{\rightarrow} \mathcal{N}\left(\mathbf{0},\boldsymbol{\gamma}(\mathbf{0})\varphi(0)\right) = \mathcal{N}\left(\mathbf{0},\frac{2}{3}\boldsymbol{\mathit{I}}\right), \quad \quad N\rightarrow \infty
\end{align*}
for every $(\boldsymbol{\mathit{x}},t)\in\mathbb{R}^4$ as $N$ tends to infinity. Due to the multi-dimensional central limit theorem, for any choice of $n\in \mathbb{N}$ and $(\boldsymbol{\mathit{x_1}},t_1), \ldots, (\boldsymbol{\mathit{x_n}},t_n) \in \mathbb{R}^4$, the joint distribution of $\tilde{\boldsymbol{\eta}}_N(\boldsymbol{\mathit{x}_1},t_1), \ldots, \tilde{\boldsymbol{\eta}}_N(\boldsymbol{\mathit{x}_n},t_n)$ converges in distribution to a normal distribution on $\mathbb{R}^{3n}$. We conclude that $\tilde{\boldsymbol{\eta}}_N$ is a centered random field with covariance \eqref{eq:cov_eta}, which is approximately Gaussian if $N$ is large. So in order to construct $\boldsymbol{\eta}$ respectively $\tilde{\boldsymbol{\eta}}_N$ we focus on the sampling of $\boldsymbol{\xi}$ and $\psi$. Thereby, we keep in mind the almost sure differentiability of the realizations of our constructed fields and processes.
 
\subsubsection{Spatial Field $\boldsymbol{\xi}$}
In this subsection we exploit the special structure of the spectral density of the spatial field given by $\boldsymbol{\mathit{S}}_{\boldsymbol{\xi}}=\mathcal{F}_{\boldsymbol{\gamma}}$ in \eqref{eq:Sloc}. Let $\boldsymbol{\mathit{w}}=(\boldsymbol{\mathit{w}}(t))_{t\in\mathbb{R}}$ be a centered, homogeneous and $\mathbb{R}^3$-valued stochastic process with spectral function 
$\boldsymbol{\mathit{S_w}}(\kappa) = E(|\kappa|)/2\,\boldsymbol{\mathit{I}}$,
i.e., its components are uncorrelated processes with the same spectral function ${s}_w(\kappa)=E(|\kappa|)/2$.
Moreover, let $\boldsymbol{\mathit{z}}$ be a uniformly distributed random vector on the unit sphere $S^2 =\{\boldsymbol{\mathit{x}}\in\mathbb{R}^3: \lVert \boldsymbol{\mathit{x}}\rVert = 1\}$. Then, under the assumption that $\boldsymbol{\mathit{w}}$ and $\boldsymbol{\mathit{z}}$ are independent, the random field
that is defined by
\begin{align*}
\boldsymbol{\xi}(\boldsymbol{\mathit{x}}) = (\boldsymbol{\mathit{I}} - \boldsymbol{\mathit{z}}\otimes \boldsymbol{\mathit{z}})\cdot\boldsymbol{\mathit{w}}(\boldsymbol{\mathit{x}}\cdot \boldsymbol{\mathit{z}})
\end{align*}
has the spectral density $\boldsymbol{\mathit{S}}_{\boldsymbol{\xi}}$ given by \eqref{eq:Sloc} and hence the desired covariance $\boldsymbol{\gamma}$, \cite{majda:p:1994, elliott:p:1995}.
Since the components of $\boldsymbol{\mathit{w}}$ are independent, it is sufficient to focus on the sampling of one component $w$ in order to construct the whole field $\boldsymbol{\xi}$. Following \cite{kurbanmuradov:p:2006}, this can be done in the subsequent manner: As $E(\kappa) \geq 0$ for all $\kappa \geq 0$ and $\int_{\mathbb{R}} s_w (\kappa)\, \mathrm{d}\kappa = \int_0^\infty E(\kappa)\,\mathrm{d}\kappa = 1$,
the function $s_w$ is a continuous probability density on $\mathbb{R}$. Choosing a random variable $R$ with this probability density and two standard normally distributed random variables $X$ and $Y$ that are all stochastically independent, the complex-valued process $(\widetilde{w}(t))_{t\in\mathbb{R}}$ 
\begin{gather*}
\widetilde{w}(t) = Z \exp(iRt), \quad Z=X+iY,
\end{gather*}
has the spectral function 2$s_w$ as a simple calculation with the conjugate-complex $\overline{\tilde w}$ shows
\begin{align*}
\mathbb{E}(\tilde{w}(t+t_1)\,\overline{\tilde w}(t_1))= \mathbb{E}\left(\exp(iRt)\right)\,\mathbb{E}(Z\overline{Z}) = 2\int_{\mathbb{R}} \exp(i\kappa t)\,s_w(\kappa)\,\mathrm{d}\kappa.
\end{align*}
By taking its real or imaginary part we obtain a real-valued process with the desired spectral function $s_w$. The so constructed process $w$ ($w = \mathrm{Re}(\widetilde{w})$ or $w = \mathrm{Im}(\widetilde{w})$) has obviously almost surely differentiable realizations and hence the same holds for $\boldsymbol{\xi}$.

\subsubsection{Time Process $\psi$}
The sampling of the time process $\psi$ can be performed analogously to $w$. We introduce the new process $\widetilde{\psi}(t) = \psi(t_{\mathrm{T}}t)$ having the covariance $\mathbb{E}(\widetilde{\psi}(t+t_1) \widetilde{\psi}(t))=\varphi(t_{\mathrm{T}} t)$, cf.\ \eqref{eq:varphi}. Consequently, its spectral function is
\begin{gather*}
s_{\widetilde{\psi}}(\omega) = \frac{1}{t_{\mathrm{T}}} \mathcal{F}_\varphi\left(\frac{\omega}{t_{\mathrm{T}}}\right)
=\frac{1}{\sqrt{2\pi}}\exp\left(-\frac{\omega^2}{2}\right).
\end{gather*}
As $s_{\widetilde{\psi}}$ is the probability density of the standard normal distribution, we take three independent, standard normally distributed random variables $R,X,Y$ and set $\widetilde{\psi}(t) = Z\exp(iRt)$ with $Z=X+iY$. Then, the process $\psi=\mathrm{Re}(\widetilde{\psi}(\cdot/{t_{\mathrm{T}}}))$ or $\psi=\mathrm{Im}(\widetilde{\psi}(\cdot/{t_{\mathrm{T}}}))$  has the desired covariance function $\varphi$ and almost surely differentiable realizations.

\quad\\
The reconstruction strategy for $\boldsymbol{\xi}$ uses the isotropic form of the spectral density and traces the construction of a random field back to the sampling of a scalar-valued stochastic process which involves an enormous reduction of complexity and effort. For the sampling of the scalar-valued process $w$ with respect to the spectral density, various (approximate) Fourier methods could be applied. However, our approach of introducing the complex-valued surrogate process via three random variables is not only exact but also efficient. The sampling and evaluation can be performed flexibly regarding the needs. In contrast to the a priori fixed discretization in the Fourier methods, this adaptivity yields advantages concerning memory and computation costs for the forthcoming simulations of the jets dynamics. The same holds true for the time process $\psi$. Here, one could certainly think of direct methods on top of the covariance, but their performance suffers from an a priori discretization and high effort (for example the effort for a Cholesky decomposition is $\mathcal{O}(n^3)$, $n$ number of grid points). Our effort is linear in the discretization. For every $\boldsymbol{\mathit{u'}}_{loc}$ we have to generate $9N$ standard normally and $3N$ $s_w$-distributed variables as well as $N$ uniformly distributed vectors on $S^2$. Thereby, the realization of the $s_w$-distributed variables is the most expensive part. These variables depend on the considered flow situation as $s_{w}(\kappa;\zeta)=E(|\kappa|;\zeta)/2$ with $\zeta=\epsilon \nu/\mathrm{k}^2$ at point $\mathrm{p}$.

\begin{remark}\label{rem:sampling}
As for the stochastic simulation (cf.\ Algorithm~\ref{alg}), a random vector $\boldsymbol{\mathit{z}}$ that is uniformly distributed on the unit sphere $S^2$  can be sampled by help of three independent standard normally distributed scalars $X_1,X_2,X_3$ according to $\boldsymbol{\mathit{z}}= (X_1,X_2,X_3)/R$ with $R=\sqrt{X_1^2 + X_2^2 + X_3^2}$ (scaling method). For the generation of a $s_w$-distributed variable we use the classical acceptance-rejection method by von Neumann \cite{neumann:p:1951}, taking the gamma distribution as reference density. For details on the techniques see e.g.\ \cite{kolonko:b:2008, mueller:b:2012}.
\end{remark}

\begin{algorithm}[Sampling Procedure]\label{alg} \quad\\
\textbf{Output}: approximate sample from $\boldsymbol{\mathit{u'}}_{loc}$ at $(\boldsymbol{x},t)$\\
\textbf{Input}: \hspace*{0.15cm} flow data at point $\mathrm{p}$: $\mathrm{k}$, $\epsilon$, $\nu$, $\mathbf{\bar u}$ and dimensionless turbulent large-scale time $t_{\mathrm{T}}$,\\
\hspace*{1.3cm} evaluation point $(\boldsymbol{x},t)$
\begin{enumerate}
\item[A.1] Determine the dimensionless local flow parameters: $\zeta=\epsilon\nu/\mathrm{k}^2$ and $\boldsymbol{\mathit{\bar u}}=\mathbf{\bar u}/\sqrt{\mathrm{k}}$
\item[A.2] Set random field parameter $N$ and generate random numbers, $l=1,\dots, N$:
\begin{itemize}
\item[$\blacktriangleright$] $\boldsymbol{z}^{(l)}$\\
$N$ samples according to the uniform distribution on the sphere $S^2$ by scaling method
\item[$\blacktriangleright$] $R_{\xi,j}^{(l)}$, $j=1,2,3$\\ 
$3N$ samples according to the density $s_w$ by von Neumann's method  \\
(obtain $s_{w}(\kappa;\zeta)=E(|\kappa|;\zeta)/2$ by solving the nonlinear system \eqref{eq:E_NGS})     
\item[$\blacktriangleright$] $X_{\xi,j}^{(l)}$, $Y_{\xi,j}^{(l)}$, $j=1,2,3$ as well as
$R_{\psi}^{(l)}$, $X_{\psi}^{(l)}$, $Y_{\psi}^{(l)}$ \\ 
$9N$ samples according to the standard normal distribution
\end{itemize}
\item[B.1] Compute approximate samples, $l=1,\dots,N$:
\begin{itemize}
\item[$\blacktriangleright$] spatial field: 
\begin{align*}
 w_j^{(l)}(\boldsymbol{x} \cdot \boldsymbol{z}^{(l)})&=\mathrm{Re}((X_{\xi,j}^{(l)}+iY_{\xi,j}^{(l)})\exp(iR_{\xi,j}^{(l)}\,\boldsymbol{x} \cdot \boldsymbol{z}^{(l)})), \quad \quad j=1,2,3\\
 \boldsymbol{\xi}^{(l)}(\boldsymbol{x}) &= (\boldsymbol{I} - \boldsymbol{z}^{(l)} \otimes \boldsymbol{z}^{(l)}) \cdot (w^{(l)}_1, w^{(l)}_2, w^{(l)}_3) (\boldsymbol{x} \cdot \boldsymbol{z}^{(l)})
 \end{align*}
\item[$\blacktriangleright$] time process: $$\psi^{(l)}(t) = \mathrm{Re}((X_\psi^{(l)} + i Y_\psi^{(l)}) \exp(i R_\psi^{(l)} t/t_{\rm T}))$$
\end{itemize}
\item[B.2] Approximate 
$$\boldsymbol{\mathit{u'}}_{loc}(\boldsymbol{\mathit{x}},t;\zeta)
\approx \frac{1}{\sqrt{N}}\sum_{l=1}^N\boldsymbol{\xi}^{(l)}(\boldsymbol{\mathit{x}}-\boldsymbol{\mathit{\bar u}}t)\,\psi^{(l)}(t)$$
\end{enumerate}
\end{algorithm}
Algorithm~\ref{alg} consists of two parts, A - the initialization with the generation of random numbers and B - the computation and evaluation. Hence,
to evaluate the same sample at a different collection of points $(\boldsymbol{x_i},t_i)$, only part B need to be executed while the initialization with the random numbers of part A should be stored.

\subsection{Simulation of $\boldsymbol{\mathit{u'}}_{loc}$ and Tests on Multivariate Normality}

\begin{table}[b]
      	\begin{tabular}{|c|c|c|c|c|c|c|c|}
        	\hline
       		$N\backslash d$ & 1 & 2 & 3 & 4 & 5 & 6\\ \hline
		10 &  0.357 & 0.416 & 0.439 & 0.461 & 0.497 & 0.518\\ \hline
		30 &  0.108 & 0.15 & 0.184 & 0.187 & 0.19 & 0.182\\ \hline
		50 &  0.098 & 0.102   & 0.12 & 0.143 & 0.183 & 0.126 \\ \hline
		70 &  0.069 & 0.079 & 0.072 & 0.087 & 0.117 & 0.124\\ \hline
		100 & 0.082 & 0.076 & 0.096 & 0.091 & 0.085 & 0.108\\ \hline
		150 & 0.094 & 0.099 & 0.093 & 0.101 & 0.109 & 0.125\\ \hline
      	\end{tabular}\\[1ex]
\caption{\label{tab:1}Rejection frequencies of Royston's test.}
\end{table}

\begin{figure}[b]\hspace*{-0.4cm}
\includegraphics[scale=0.395]{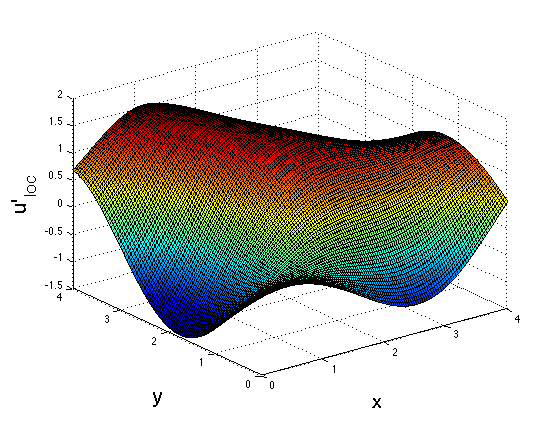}\hspace*{-0.6cm}
\includegraphics[scale=0.395]{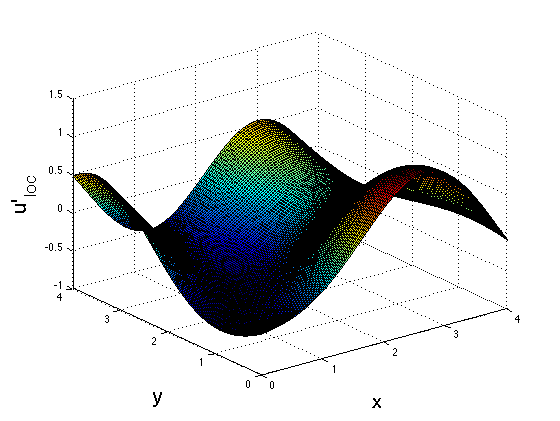}
\caption{\label{fig:1} Realization of a component of $\boldsymbol{\mathit{u}'}_{loc}$ with $\zeta=0$, plotted over two-dimensional space at certain times.}
\end{figure}

Simulating the local velocity fluctuations, the finite-dimensional distributions of $\boldsymbol{\tilde{\eta}}_N$ are close to a multivariate normal distribution for large $N$ according to the central limit theorem. For the assessment of the multivariate normality on a fixed set of points $\{(\mathbf{x_1},t_1),\dots,(\mathbf{x_d},t_d)\}\subset\mathbb{R}^3\times\mathbb{R}_0^+$ we apply the statistical test by Royston \cite{royston:p:1982}, \cite{royston:p:1983}  which we evaluate by help of the respective MATLAB routine \cite{trujillo:w:2007a}. Table~\ref{tab:1} shows the rejection frequencies at the significance level 0.05 for different values of the variate size $d$ and the random field parameter $N$. We use here 1000 Monte Carlo replications and a sample size of 50. The rejection frequency among the 1000 replications turns out to be a robust quantity that is widely independent of the variate size $d$, moreover it stays approximately the same for $N\geq 50$. Hence, we use $N=50$ in \eqref{eq:Gauss} for the forthcoming simulations. The observed rejection frequency of 10\% is acceptable for us since the Gaussian distribution is a secondary property. Our main interest is the accurate construction of the covariance structure. Figure~\ref{fig:1} illustrates the numerical result of a realization of $\boldsymbol{\mathit{u}'}_{loc}$ for a typical flow situation with $\zeta=0$.

\subsection{Globalization Strategy -- Realization of the Global Velocity Fluctuations}

According to the Global-from-Local-Assumption (Assumption~\ref{ass:1}), the global velocity fluctuations $\mathbf{u'}(\mathbf{x},\mathrm{t})$ can be obtained from the Gaussian average over the local information of the fields $\mathbf{u'}_{loc,\mathrm{p}}$ that belong to the neighborhood $\mathrm{p}\in M(\mathbf{x},\mathrm{t})$ of relevant space and time correlations \eqref{eq:u'}. This construction satisfies the requirements of the $\mathrm{k}$-$\epsilon$ turbulence model \eqref{eq:keps} in an integral sense, \cite{marheineke:p:2006}. It becomes already very memory-demanding and time-consuming for a discretization with $|M(\mathbf{x},\mathrm{t})|\geq 5$ which is obviously rather coarse in $\mathbb{R}^4$. In view of a turbulent spinning process it is not applicable. 

Thus, we propose the following globalization strategy
\begin{align}\label{eq:global} 
\mathbf{u'}(\mathbf{x},\mathrm{t})=\mathrm{k}^{1/2}(\mathrm{p})\,\,\,\boldsymbol{\mathit{u'}}_{loc,\mathrm{p}}\left(\frac{\epsilon}{\mathrm{k}^{3/2}}(\mathrm{p}) \,\mathbf{x},\frac{\epsilon}{\mathrm{k}}(\mathrm{p})\,\mathrm{t}; \frac{\epsilon \nu}{\mathrm{k}^2}(\mathrm{p}) \right)\Big|_{\mathrm{p}=(\mathbf{x}, \mathrm{t})}
\end{align}
using the space- and time-dependent flow fields for kinetic turbulent energy, dissipation rate and viscosity known from the $\mathrm{k}$-$\epsilon$-simulation.
The global velocity fluctuation field of \eqref{eq:global} fulfills the condition \eqref{eq:keps} on the kinetic turbulent energy exactly, i.e.\ $\mathbb{E}(\mathbf{u}'(\mathbf{x},\mathrm{t})\cdot \mathbf{u}'(\mathbf{x},\mathrm{t}))=2\mathrm{k}(\mathbf{x},\mathrm{t})$ for all $(\mathbf{x},\mathrm{t})\in \mathbb{R}^3\times \mathbb{R}_0^+$. Furthermore, the condition on the dissipation rate containing the spatial derivatives is valid up to an error of order $\mathcal{O}(\zeta_0)$ with constant $\zeta_0=\epsilon_0 \nu_0/\mathrm{k}^2_0 \ll 1$ representing the typical ratio of turbulent fine-scale and large-scale length. This estimate is based on the observation that changes in the behavior of $\mathrm{k}$ and $\epsilon$ mainly appear on the large scale and not on the fine scale. This motivates an asymptotic consideration with the multi-scale ansatz $\mathrm{k}(\mathbf{x},\mathrm{t})=\mathrm{k_0}+\mathrm{k_1}(\zeta_0 \mathbf x, \mathrm{t})$ (and analogously for $\epsilon$, $\nu$), yielding the result.

Since the whole field $\zeta(\mathbf{x},\mathrm{t})=\epsilon \nu/\mathrm{k}^2 (\mathbf{x},\mathrm{t})$ is negligibly small in turbulent flows (see e.g.\ Figure~\ref{fig:scales} for a turbulent air stream in a melt-blowing process), \eqref{eq:global} might be further simplified to 
\begin{align}\label{eq:global1}
\mathbf{u'}(\mathbf{x},\mathrm{t})=\mathrm{k}^{1/2}(\mathrm{p})\,\,\,\boldsymbol{\mathit{u'}}_{loc,\mathrm{p}}\left(\frac{\epsilon}{\mathrm{k}^{3/2}}(\mathrm{p}) \,\mathbf{x},\frac{\epsilon}{\mathrm{k}}(\mathrm{p})\,\mathrm{t}; 0) \right)\Big|_{\mathrm{p}=(\mathbf{x}, \mathrm{t})}.
\end{align}
Considering an asymptotic expansion in $\zeta_0$, \eqref{eq:global1} and \eqref{eq:global} obviously agree in leading order.
By this slight modification, the sampling of the global velocity fluctuations $\mathbf{u'}$ in \eqref{eq:global1} can be performed with respect to a parameter-free energy spectrum $E(\,\cdot\,;\zeta=0)$ \eqref{eq:E}, which avoids the expensive solving of different nonlinear systems \eqref{eq:E_NGS} and the multiple application of von Neumann's method for all the required $\boldsymbol{\mathit{u'}}_{loc,\mathrm{p}}$ (see Algorithm~\ref{alg}, Step A.2). So, in total we only need a single set of random parameters ($3N$ $s_w$-distributed variables with $s_w(\kappa;0)=E(|\kappa|;0)/2$, $\kappa\in\mathbb{R}$ as well as $9N \sim \mathcal{N}(0,1)$ and $N$ uniformly distributed vectors on $S^2$). This involves an enormous decrease of computational costs and makes the globalization strategy applicable to practically relevant problems as we will show. Note that its realization is straightforward based on Algorithm~\ref{alg}.

\section{Application to a Turbulent Spinning Process}\label{sec:4}

\subsection{Simplified ODE-Model}

The characteristics of a turbulent spinning process like melt-blowing are the huge jet elongations (jet thinning) that are obtained by the stretching due to turbulent air flows. Up to now, the numerical simulations available in the literature, e.g.\ \cite{uyttendaele:p:1990, zeng:p:2011, xie:p:2012}, cannot predict the large elongations measured in the experiments $e \sim \mathcal{O}(10^6)$. We suppose that the reason lies in the steady considerations of the turbulent flow field and the neglect of the fluctuations. In \cite{sinha-ray:p:2010} perturbations (bending instability) on a jet were imposed by turbulent pulsations, leading to stretching and thinning. In \cite{marheineke:p:2006, marheineke:p:2011} we developed a model framework for the dynamics of a long slender object (fiber) in a turbulent flow in terms of a random aerodynamic drag force in a one-way-coupling. The dimensionless drag force $\boldsymbol{\mathit{f}}$ depends on the relative velocity between air flow and object as well as on the object's tangent. To study the impact of our random force model and to get first qualitative estimates for the jet attenuation, we deal here with a simplified model for an isothermal fiber jet dynamics. Instead of the complex PDE-based Cosserat models that contain inner stresses and temperature dependencies \cite{antman:b:2006, arne:p:2010, arne:p:2011}, it just consists of a system of first order ODEs in time for jet position $\mathbf{r}$, velocity $\mathbf{v}$ and elongation $e$. For this purpose, we approximate the jet tangent (space derivative) in the drag force $\boldsymbol{\mathit{f}}$ by the direction of the jet velocity $\mathbf{v}/\|\mathbf{v}\|$ and motivate an evolution equation for the elongation from the steady situation where $e=\|\mathbf{v}\|/\mathrm{v}_0$ with exit velocity magnitude at the spinning nozzle $\mathrm{v}_0$ (cf.\ Section~\ref{sec:1}). The resulting model \eqref{eq:ODE} describes the path and behavior of a single jet point whose motion is exclusively driven by a turbulent air flow. Certainly, gravitational forces could be included, but they play a negligibly small role for the observed jet thinning. The constructed Gaussian field $\mathbf{u}=\bar{\mathbf{u}}+\mathbf{u}'$ for the turbulent flow velocity carries the randomness into the dynamic system via the drag force,
\begin{align}\label{eq:ODE}
\frac{\mathrm{d}}{\mathrm{dt}}\mathbf{r} &= \mathbf{v} 
&& \mathbf{r}(0) = \mathbf{r_0}\\ \nonumber
\frac{\mathrm{d}}{\mathrm{dt}}\mathbf{v} &= e^{3/2} \, \mathrm{a}(\mathbf{r},\mathrm{t}) \,
\boldsymbol{\mathit{f}}\left(\frac{\mathbf{v}}{\|\mathbf{v}\|},\frac{1}{\sqrt{e}}
\frac{\mathbf{u}(\mathbf{r},\mathrm{t})-\mathbf{v}}{\mathrm{b}(\mathbf{r},\mathrm{t})}\right)
&& \mathbf{v}(0) = \mathrm{v}_0\boldsymbol{\tau_0}\\ \nonumber
\frac{\mathrm{d}}{\mathrm{dt}}e &= \frac{1}{\mathrm{v}_0}\, e^{3/2} \, \mathrm{a}(\mathbf{r},\mathrm{t})
\,\, \left\|\boldsymbol{\mathit{f}}\left(\frac{\mathbf{v}}{\|\mathbf{v}\|},\frac{1}{\sqrt{e}}
\frac{\mathbf{u}(\mathbf{r},\mathrm{t})-\mathbf{v}}{\mathrm{b}(\mathbf{r},\mathrm{t})}\right)\right\|
&& e(0) = 1
\end{align}
where the scalar-valued functions $\mathrm{a}$ and $\mathrm{b}$ 
\begin{align*}
\mathrm{a} = \frac{4}{\pi}\frac{\rho\nu^2}{\rho_{\mathrm{F}} \mathrm{d}_0^3},\quad \quad \quad \mathrm{b} = \frac{\nu}{\mathrm{d}_0}
\end{align*}
contain -- apart from constant fiber jet quantities ($\rho_{\mathrm{F}}$ density, $\mathrm{d}_0$ diameter at the nozzle) -- also space- and time-dependent air flow information ($\rho$ density, $\nu$ viscosity).

\begin{remark}[Numerical Treatment]\label{rem:1}
For the forthcoming numerical investigations of the random ordinary differential system \eqref{eq:ODE} the $\mathrm{k}$-$\epsilon$-simulations of the underlying turbulent flow field are performed with the software ANSYS Fluent. The fiber jet dynamics is computed in MATLAB using the standard ODE-solver \texttt{ode45.m}. The routine is an implementation of an explicit Runge-Kutta method of fourth (or respectively, fifth) order with adaptive time step control. The random normally and uniformly distributed numbers that are required for the sampling of the turbulent velocity fluctuations $\mathbf{u'}$ are generated with the MATLAB-functions \texttt{randn()} and \texttt{rand()}.
\end{remark}

\subsection{Numerical Results}

\begin{figure}[H]
\includegraphics[scale=0.34]{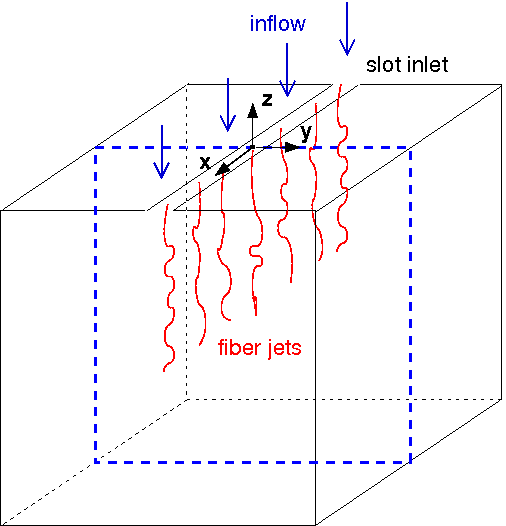} \hspace*{0.5cm}
\includegraphics[scale=0.34]{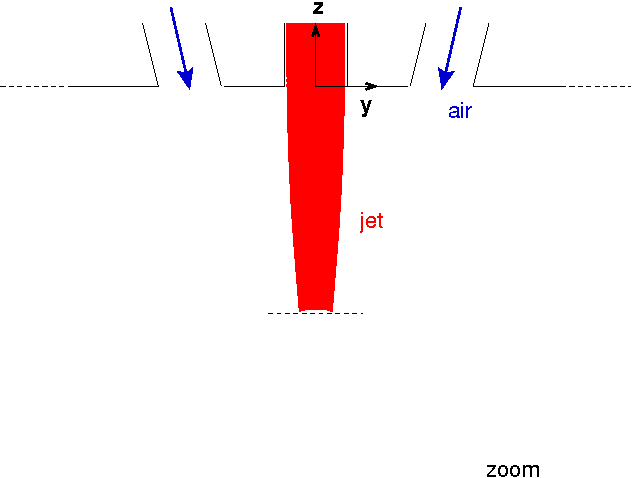}
\caption{\label{fig:sketch} Sketch of flow domain with immersed fiber jets. A two-dimensional cut ($\mathrm{y}$-$\mathrm{z}$-plane, marked by dashed line) is representative due to the given homogeneity in $\mathrm{x}$-direction. For details on possible geometries see e.g.\ \cite{malkan:p:1995, pinchuk:b:2002}.}
\end{figure}

\begin{figure}[H]\hspace*{-0.75cm}
\includegraphics[scale=0.4]{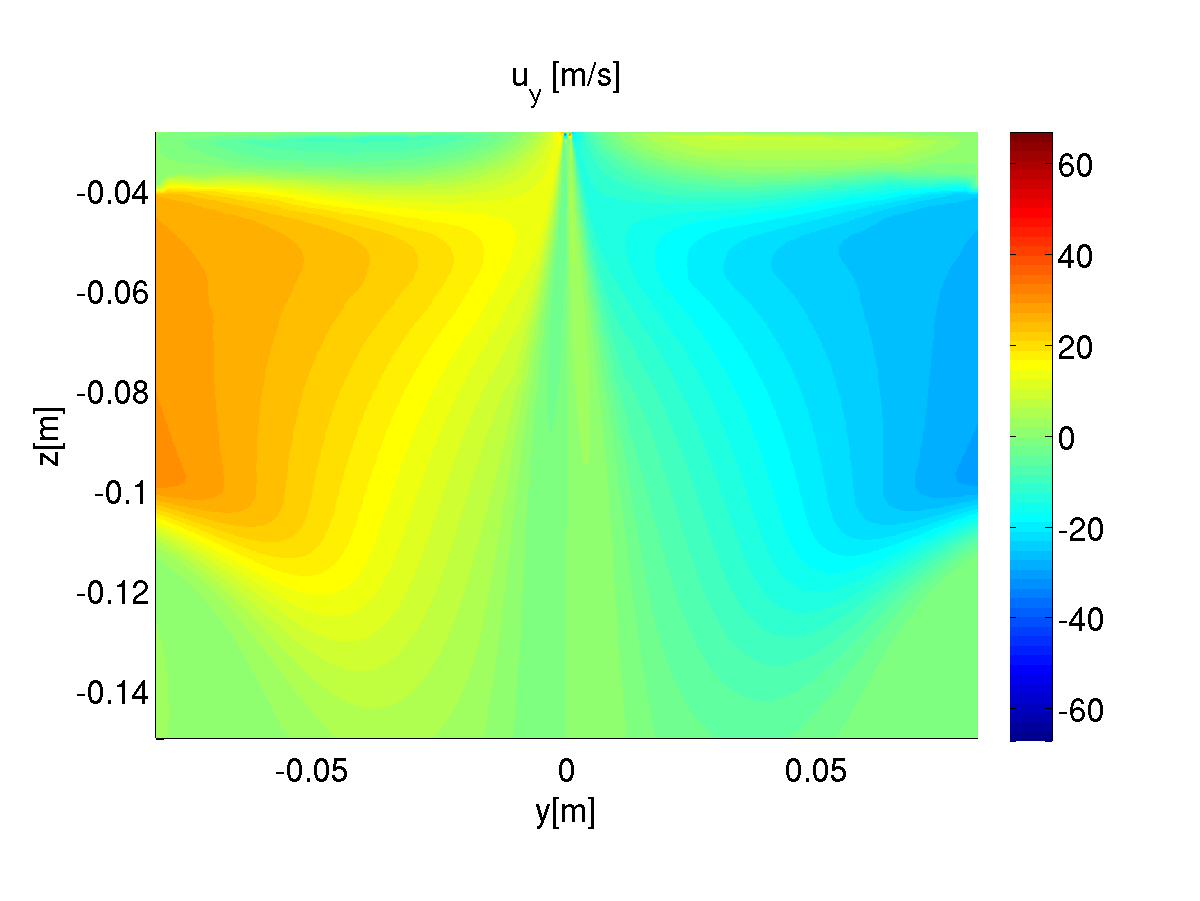} \hspace*{-0.75cm}
\includegraphics[scale=0.4]{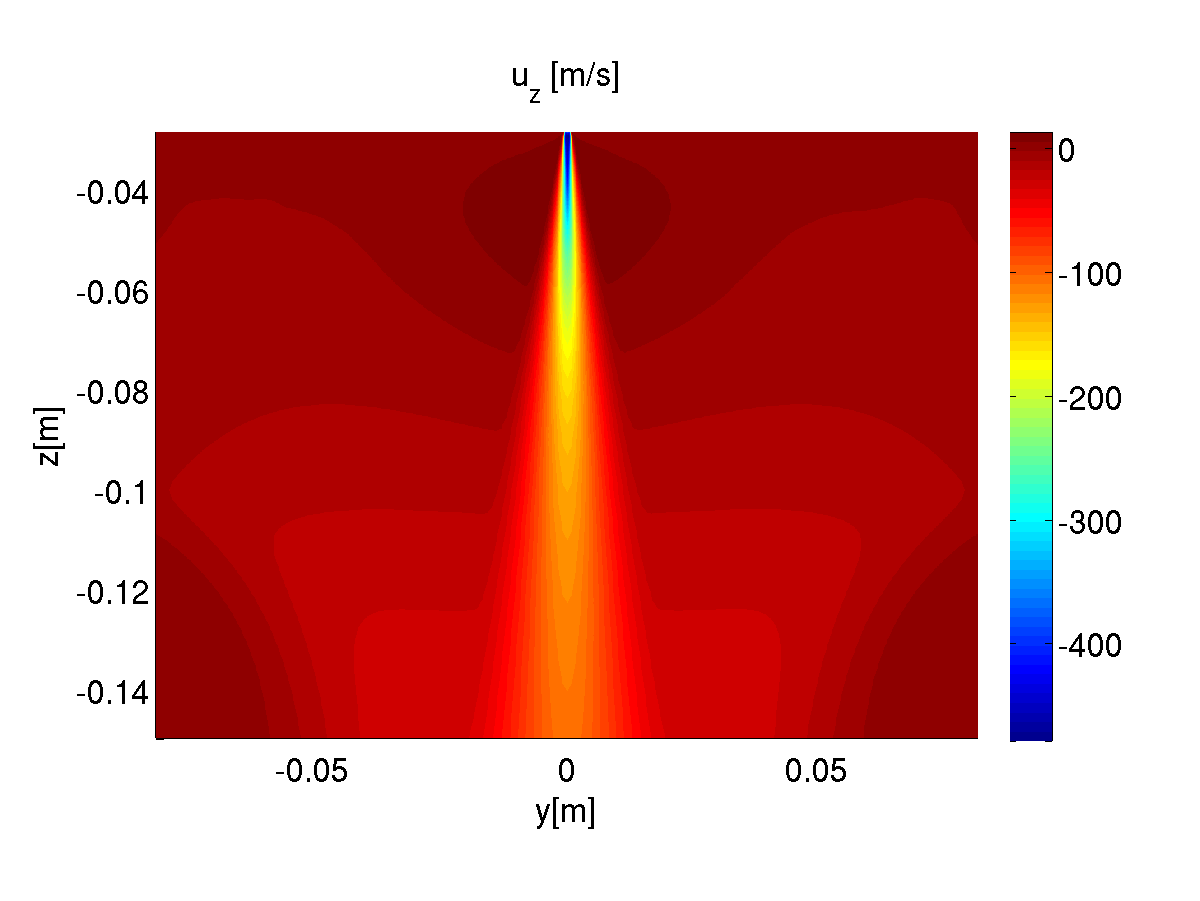}\\\vspace*{-0.5cm}\hspace*{-0.75cm}
\includegraphics[scale=0.4]{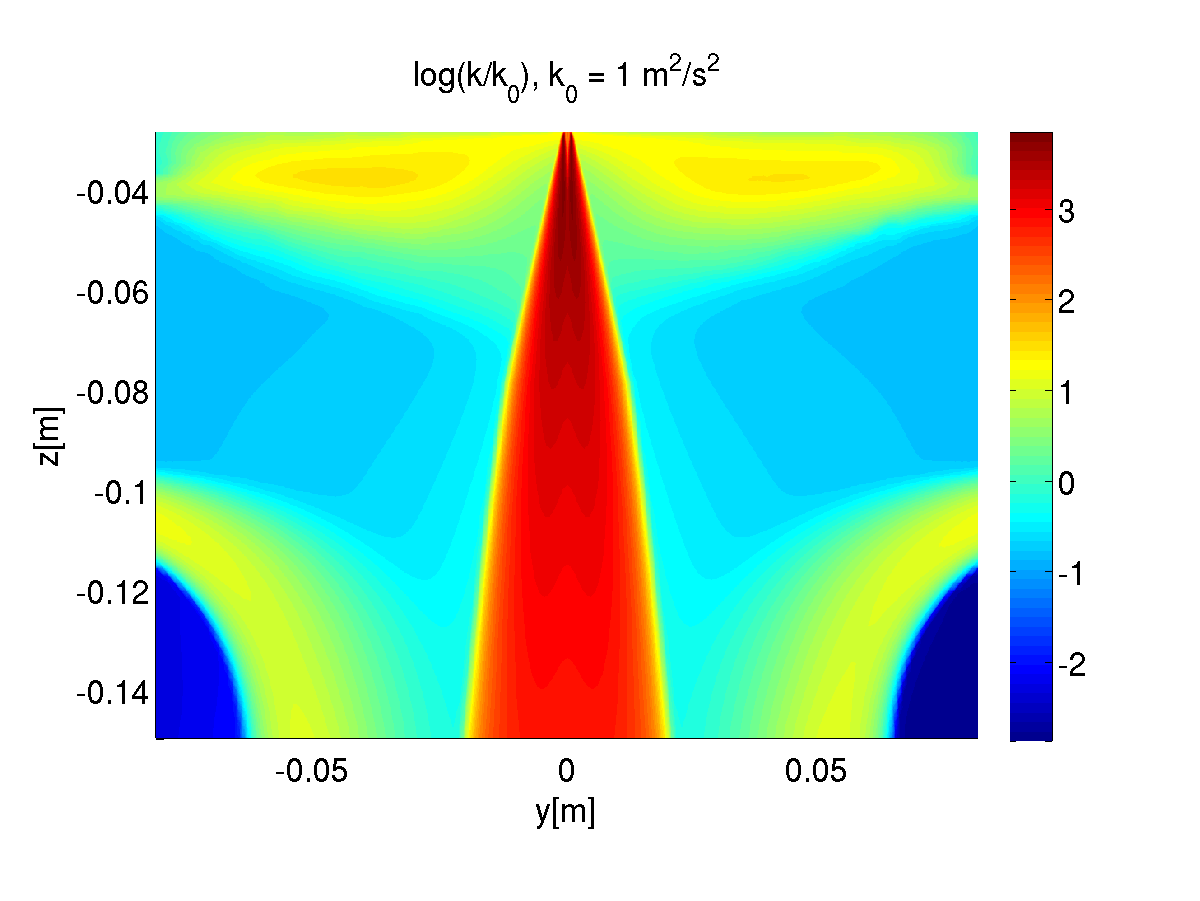} \hspace*{-0.75cm}
\includegraphics[scale=0.4]{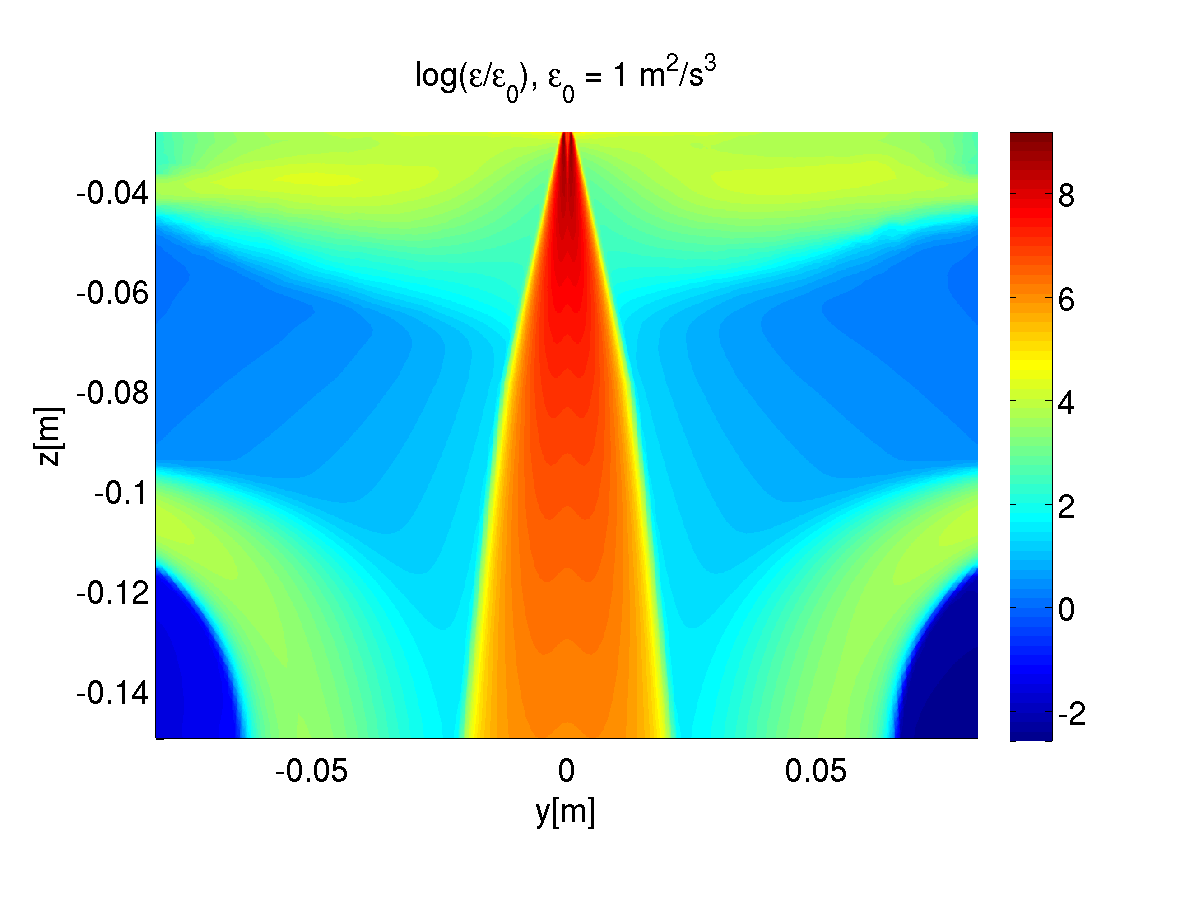}\vspace*{-0.5cm}
\caption{\label{fig:keps-simulation}$\mathrm{k}$-$\epsilon$-simulation of the representative 2d flow domain. \textit{Top:} components of mean velocity $\mathbf{\bar u}$ in $\mathrm{y}$- and $\mathrm{z}$-direction. \textit{Bottom:} turbulent kinetic energy $\mathrm{k}$ and dissipation rate $\epsilon$ in logarithmic plots.}
\end{figure}

\begin{figure}[H]
\hspace*{-0.8cm}
\includegraphics[scale=0.4]{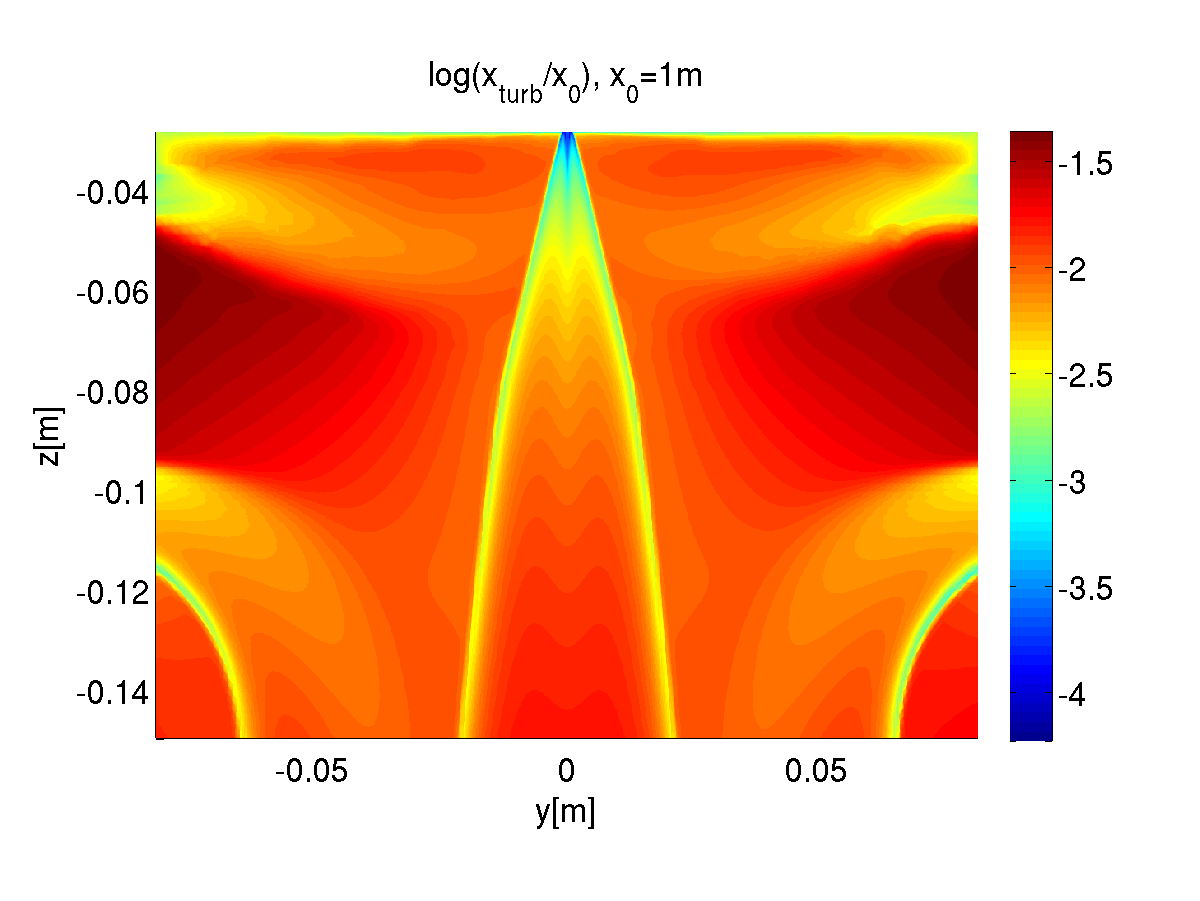}\hspace*{-0.6cm}
\includegraphics[scale=0.4]{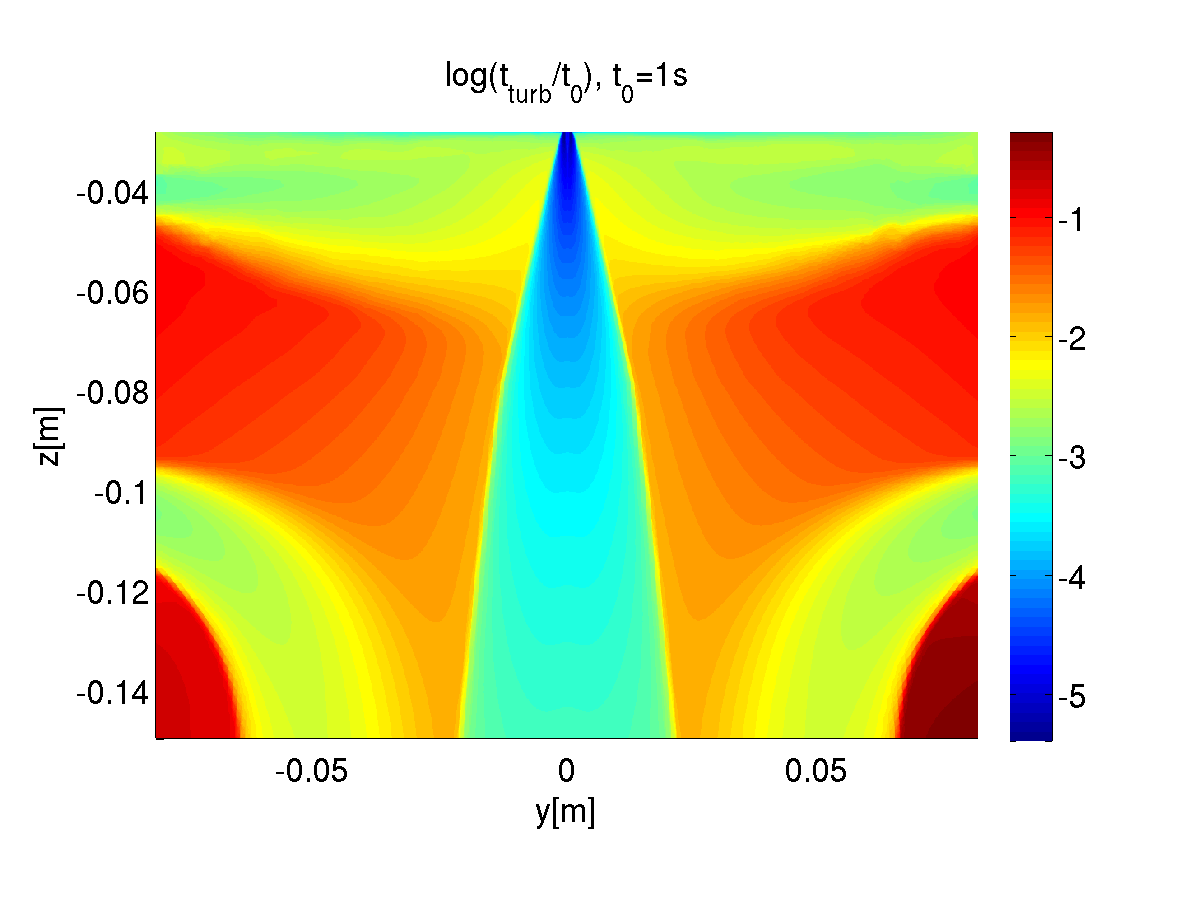}\\\vspace*{-0.5cm}
\includegraphics[scale=0.4]{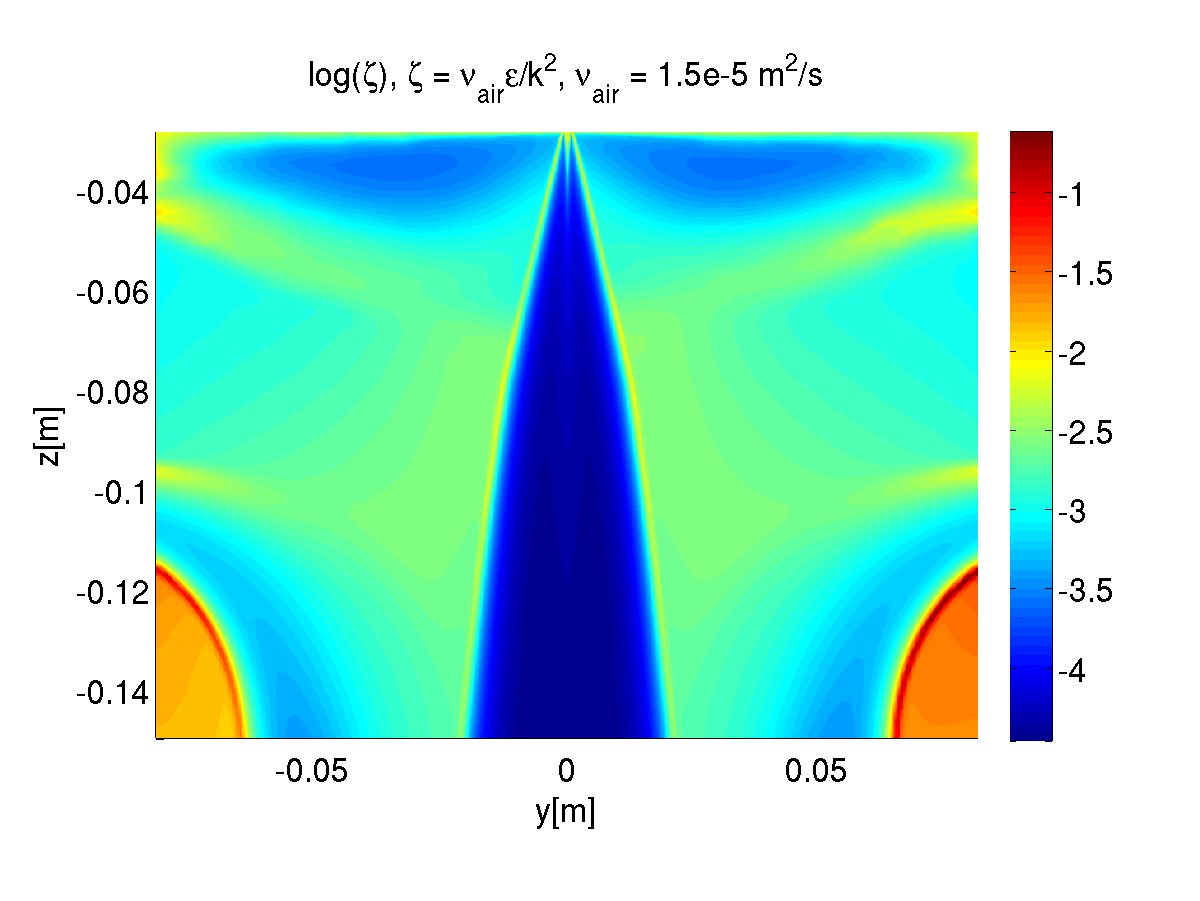}\vspace*{-0.5cm}
\caption{\label{fig:scales}Turbulent scales corresponding to Figure~\ref{fig:keps-simulation} in logarithmic plots. \textit{Top:} turbulent large-scale length $\mathrm{l_T}=\mathrm{k}^{3/2}/\epsilon$ and time $\mathrm{t_T}=\mathrm{k}/\epsilon$. \textit{Bottom}: ratio of fine and large scales $\zeta=\epsilon \nu/\mathrm{k}^2$.}
\end{figure}

We investigate the dynamics and behavior of a fiber jet in a flow situation that is usual for melt-blowing, \cite{malkan:p:1995, pinchuk:b:2002}. Temperature effects are neglected for simplicity. The air stream is directed vertically downwards and enters the domain of interest via a thin slot die, cf.\ Figure~\ref{fig:sketch}. Since the mean quantities of the turbulent air stream are time-independent and homogenous in direction of the slot ($\mathrm{x}$-direction), we perform stationary $\mathrm{k}$-$\epsilon$-simulations for a representative two-dimensional cut showing the $\mathrm{y}$-$\mathrm{z}$-plane. In the set-up the mean flow is sym\-metric with respect to the $\mathrm{z}$-axis ($\mathrm{y}=0$). Figure~\ref{fig:keps-simulation} shows the respective flow fields for the mean velocity components, kinetic turbulent energy and dissipation rate; in addition $\rho \approx 1$ [kg/m$^3$] and $\nu=1.5\cdot10^{-5}$ [m$^2$/s] is constant. A distinct free air jet develops that is supersonic at the inlet slot (here: $\|\mathbf{\bar{u}}\|\approx 400$ [m/s], $\mathrm{k}\approx 10^{3}$ [m$^2$/s$^2$], $\epsilon\approx 10^{8}$ [m$^2$/s$^3$]) and becomes subsonic within some centimeters away. So, the occurring typical turbulent length and time scales lie in a wide range, i.e., $\mathrm{l_T}=\mathrm{k}^{3/2}/\epsilon \in (10^{-4},10^{-2})$ [m] and $\mathrm{t_T}=\mathrm{k}/\epsilon \in (10^{-5},10^{-3})$ [s]. But, $\zeta\approx 10^{-4}$ in the whole free air stream, as visualized in Figure~\ref{fig:scales}. 
At the boundaries of the flow domain we observe side effects coming from the geometry (e.g.\ in the lower corners). However, these play no role for the dynamics of the fiber jet. Consequently, the simplified globalization strategy (asymptotic limit $\zeta = 0$ \eqref{eq:global1}) for the sampling of $\mathbf{u'}$ is acceptable and applied.

\newpage

\begin{figure}[H]
\hspace*{-0.2cm}
\includegraphics[scale=0.4275]{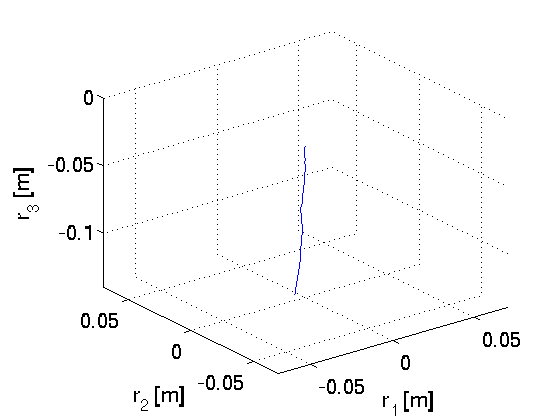}\hspace*{-0.4cm}
\includegraphics[scale=0.425]{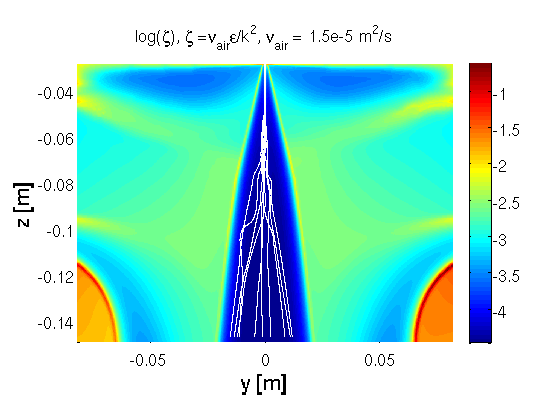}
\vspace*{-0.6cm}      
\caption{\label{fig:jet}Fiber dynamics driven by turbulent aerodynamic drag force. \textit{Left:} random trajectory $\mathbf{r}$. \textit{Right:} projection of several trajectories into $\mathrm{y}$-$\mathrm{z}$-plane (white curves). They are located in the distinct free air stream where $\zeta\approx 10^{-4}\ll 1$.}
\end{figure}

\begin{figure}[H]
\hspace*{-0.2cm}
\includegraphics[scale=0.4]{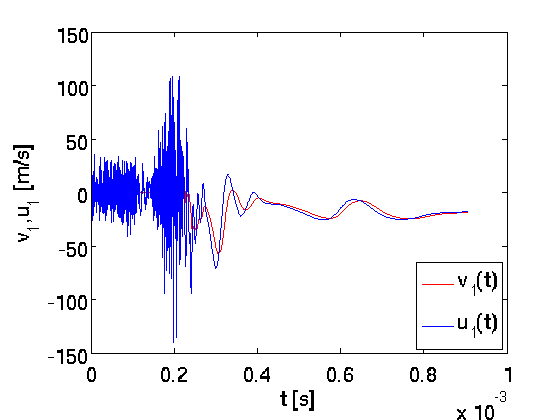} \hspace*{-0.3cm}
\includegraphics[scale=0.4]{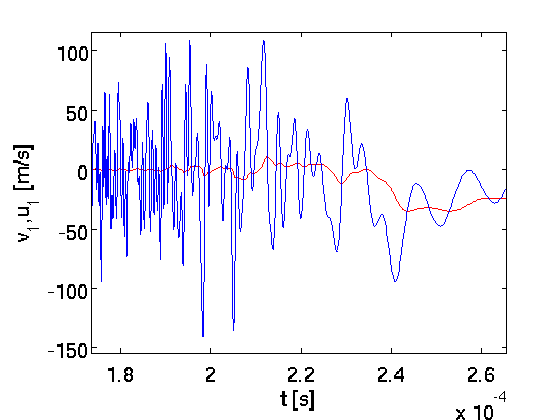}\\
\hspace*{-0.2cm}
\includegraphics[scale=0.4]{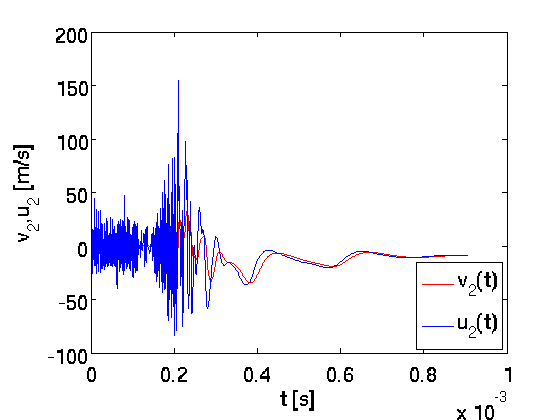} \hspace*{-0.3cm}
\includegraphics[scale=0.4]{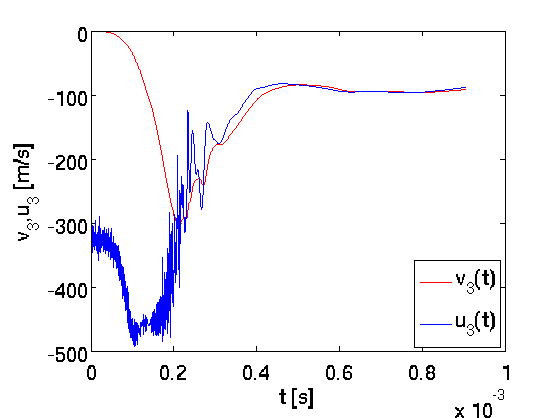}     
\vspace*{-0.3cm}                          
\caption{\label{fig:velocity}Temporal evolution of fiber $\mathbf{v}$ and air flow velocity $\mathbf{u}=\mathbf{\bar u}+\mathbf{u'}$ experienced by the random trajectory of Fig.~\ref{fig:jet} (left); component-wise visualization. The plot \textit{top-right} is a zoom in the $\mathrm{v}_1$, $\mathrm{u}_1$-components ($\mathrm{x}$-direction).}
\end{figure}

\begin{figure}[t]\hspace*{-0.2cm}
\includegraphics[scale=0.475]{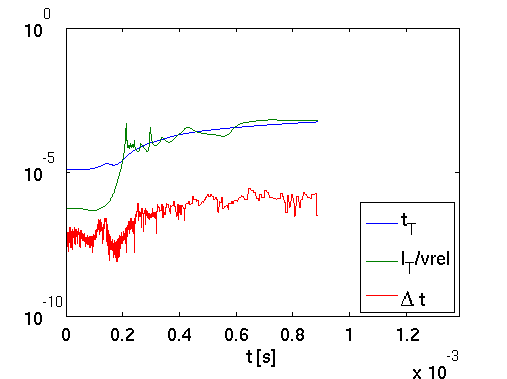}
\caption{\label{fig:resolution} Adaptive time step choice $\Delta \mathrm{t}$ for \eqref{eq:ODE} in comparison to turbulent time scales $\mathrm{t_T}=\mathrm{k}/\epsilon$ and $\mathrm{l_T}/\mathrm{v}_{rel}=\mathrm{k}^{3/2}/(\epsilon \mathrm{v}_{rel})$ with $\mathrm{v}_{rel}=\|\mathbf{\bar u}-\mathbf{v}\|$ experienced at the jet position $\mathbf{r}(\mathrm{t})$ of Fig.~\ref{fig:jet}.}
\end{figure}

The immersed fiber jet to be spun in (negative) $\mathrm{z}$-direction is initialized at the slot die (spinning nozzle) with $\mathrm{v}_0=10^{-2}$ [m/s], $\mathrm{d}_0=4\cdot 10^{-4}$ [m] and $\rho_{\mathrm F}=7\cdot10^2$ [kg/m$^3$] and simulated for the time interval $[0,\mathrm{T})$, $\mathrm{T}=10^{-3}$ [s]. The fiber jet moves exclusively in the distinct region of the free air stream, Figure~\ref{fig:jet}. Thereby, its motion is determined strongly by the mean flow pulling the fiber jet straight downwards, on the one hand. On the other hand the flow fluctuations cause a slight bouncing. The fiber velocity follows and finally adjusts to the flow velocity, as we can see in Figure~\ref{fig:velocity}. In the temporal evolution the fiber point starts from the nozzle where the impact of the turbulence is at the strongest. The velocity fluctuations act here on tiny length and time scales causing a quick acceleration and a very strong stretching of the jet. When the fiber point is some centimeters away from the nozzle after 0.2-0.3 milliseconds, the turbulence attenuates and the turbulent scales become larger (Figures~\ref{fig:scales} and \ref{fig:resolution}). In particular, $\mathrm{t_T}$ and the jet's reaction time coincide which can be concluded from the velocity curves that match. Also the elongation stagnates. 

\newpage

\begin{figure}[H]
\hspace*{-0.2cm}
\includegraphics[scale=0.47]{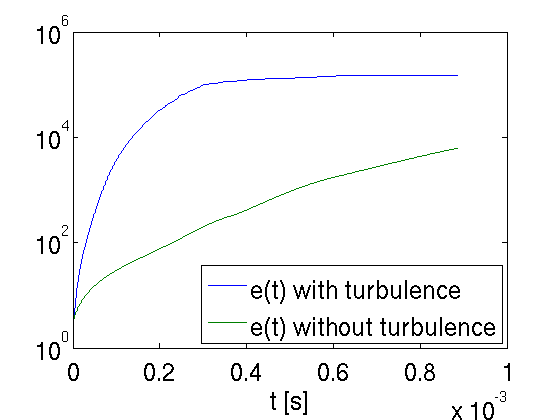}
\includegraphics[scale=0.375]{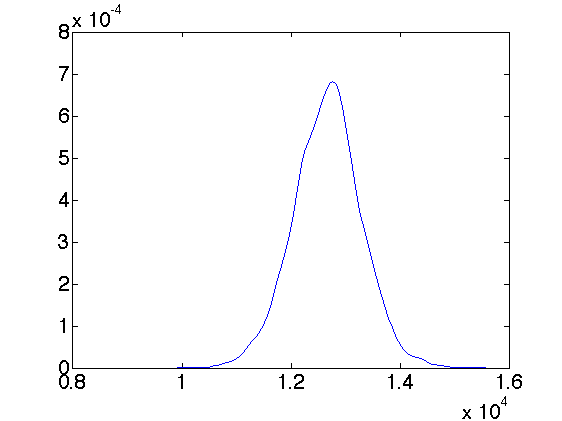}\\
\vspace*{-0.35cm}
\footnotesize{\hspace*{7cm}a) $\mathrm{r_3}=-0.033$\,[m] ($\mathrm{t}\approx 1.5\cdot10^{-4}$\,[s])}\\
\includegraphics[scale=0.375]{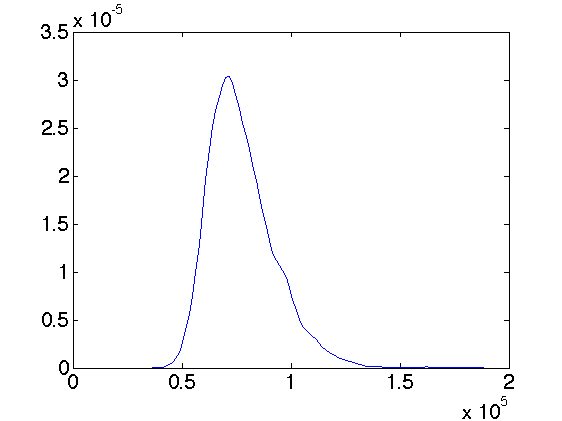}
\includegraphics[scale=0.375]{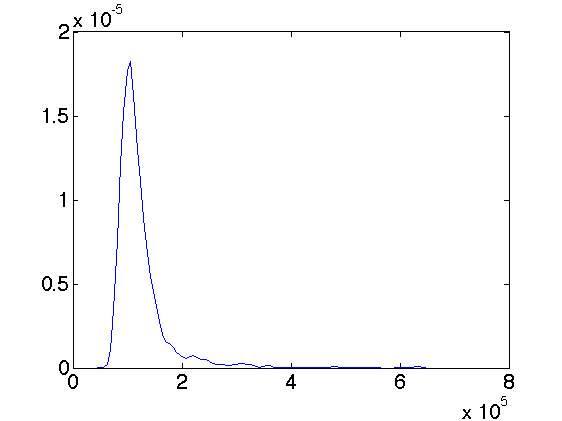}\\
\vspace*{-0.35cm}
\footnotesize{\hspace*{-0.95cm}b) $\mathrm{r_3}=-0.066$\,[m] ($\mathrm{t}\approx 2.9\cdot10^{-4}$\,[s]) \hspace*{2.6cm} c) $\mathrm{r_3}=-0.1$\,[m] ($\mathrm{t}\approx 5.4\cdot10^{-4}$\,[s])}\\
\caption{\label{fig:elongation} Elongation. \textit{Top left:} (sample) fiber elongation $e$ over time with regard and neglect of turbulent velocity fluctuations $\mathbf{u'}$. \textit{Other plots a)-c):} probability density of $e$ at depicted heights $\mathrm{r_3}=-0.033, -0.066, -0.1$ [m] ($\mathrm{z}$-direction) estimated with Monte-Carlo simulation.}
\end{figure}

Figure~\ref{fig:elongation} shows the probability density function of the elongation at three depicted heights $\mathrm{r_3}=-0.033,$ $-0.066$, $-0.1$ [m]. It is estimated by help of Monte-Carlo simulations with 5000 samples. We clearly observe the essential effect of the turbulent velocity fluctuations and our random aerodynamic drag force model on the jet thinning (especially in the first centimeters / tenths of milliseconds). The computed elongation rises up to a mean of $2\cdot 10^{5}$ at $\mathrm{T}$ (here: $\mathrm{r_3}\approx -0.14$ [m]). In comparison, the numerical result neglecting the fluctuations is approximately $10^{4}$ which perfectly corresponds to the theoretical considerations on stationary turbulence stating that $e=\|\mathbf{v}\|/\mathrm{v}_0$ with $\mathbf{v}\approx \mathbf{\bar u}$ holds, see Figure~\ref{fig:elongation}. The fact that Zeng et al.\ \cite{zeng:p:2011} have obtained the same magnitude of elongation for an visco-elastic spring-beam jet model in a mean turbulent flow field clearly stresses that the turbulent fluctuations are the major dominant effect for the large jet attenuation; material models and inner stresses in contrast seem to be of minor relevance. Moreover, it is worth to mention, that our ODE-model -- as simple as it is -- already predicts qualitatively appropriately all jet thinning stages observed in the experiments. However, proper quantitative estimates can be only expected according to the measurements \cite{bansal:p:1998} when temperature dependencies (e.g.\ temperature-dependent viscosity) are included. 

Summing up, our numerical results are very promising. They raise hope that our proposed approach with the random aerodynamic drag force is capable of predicting the large elongations that are measured in industrial melt-blowing processes, presupposing an appropriate Cosserat model for the viscous, non-isothermal fiber jet. In addition, the computational effort seems to be manageable since the asymptotic globalization strategy for the sampling of the turbulent velocity fluctuations is linear in time and space discretization.

\begin{remark}
Some concluding remarks on computational aspects: 
\begin{itemize}
\item The adaptive time step control of the ODE-solver (cf.\ Remark~\ref{rem:1}) ensures the correct resolution of the turbulent scales since the chosen step size $\Delta \mathrm{t}$ is always clearly smaller than $\mathrm{t_T}$ and $\mathrm{l_T}/\mathrm{v}_{rel}$, $\mathrm{v}_{rel}=\|\mathbf{\bar u}-\mathbf{v}\|$, cf.\ Figure~\ref{fig:resolution}. This implies the smooth numerical approximation of the jet quantities, see e.g.\ the visualization of the velocity components in Figure~\ref{fig:velocity}. 
\item The simulation of a fiber trajectory on $[0,\mathrm{T})$, $\mathrm{T}=10^{-3}$ [s] takes a CPU-time of approximately $200$ seconds on a 2.7 GHz Intel Core i5 processor.
\item The computational effort of the sampling routine for the turbulent velocity fluctuations $\mathbf{u'}$ splits into initialization and continuous run. Whereas the costs for the initial generation of the set of random numbers are independent of the discretization and negligibly small (0.1 CPU-seconds for $N=50$), the costs for the continuous run are linear in the time discretization and add up to approximately 88\% of the total costs for solving the random ODE system \eqref{eq:ODE}. On the first glance this seems to be incredibly much but the reason lies in the necessary processing of the underlying flow data (e.g.\ sorting, interpolation of flow data are required). So far, no further attention has been paid to the data processing that is done with standard MATLAB routines. But its performance will be optimized in future which promises a drastical speed-up. 
\end{itemize}
\end{remark}

\section{Conclusion and Outlook}

In a melt-blowing process liquid fiber jets are spun due to turbulent air streams causing very high jet attenuation and final diameters of size smaller than a micrometer. So far, the understanding and design of the process suffered from a discrepancy between measurements and simulations; the computed final jet diameters were too thick by several orders of magnitude. This gap might be closed by considering the impact of the turbulent velocity fluctuations on the jets dynamic as we have demonstrated numerically in this paper. In correspondence to turbulence theory and on top of a $\mathrm{k}$-$\epsilon$ formulation we have modeled the turbulent velocity fluctuations as Gaussian random fields. Taking advantage of the special covariance/correlation structure we have proposed a fast and accurate sampling strategy whose effort is linear in the discretization and that makes the realization possible. The numerical results are very convincing as they show already a qualitatively appropriate jet thinning in magnitude for a quite simple isothermal ODE-model for the jet dynamics. 

In future we intend to apply the developed random aerodynamic drag force to more sophisticated Cosserat models including also inner stresses and temperature dependencies in order to get quantitative estimates for melt-blowing. But this requires the robust numerical treatment of the PDE-models \cite{arne:p:2010, arne:p:2012, audoly:p:2012}. Especially, the handling of the expected huge elongations pose severe challenges on an efficient adaptive mesh refinement which is topic of recent research.

\quad\\
{\sc Acknowledgments.} Special thanks go to the Department of Transport Processes, Fraunhofer ITWM for the air flow simulations of the melt-blowing process. This work has been supported by German Bundesministerium f\"ur Bildung und Forschung, Schwerpunkt "Mathematik f\"ur Innovationen in Industrie und Dienstleistungen", Projekt 03MS606, and the Fraunhofer Innovationszentrum "Applied System Modeling", Kaiserslautern. 


\end{document}